\documentclass[twocolumn]{article}
\usepackage{amsfonts}
\usepackage[a4paper, total={6in, 8in}]{geometry}
\usepackage{bm}
\usepackage{amsmath}
\usepackage{hyperref}
\usepackage{multicol}
\usepackage{amsfonts}
\newcommand{\mf}{\mathcal{F}}

\newcommand{\vo}{v_{ex}^0}
\usepackage{graphicx}
\usepackage{amsthm}
\usepackage{cleveref}
\usepackage{subcaption}
\usepackage{authblk}
\usepackage{fullpage}
\usepackage{xcolor}
\usepackage{float}
\usepackage{cuted}
\DeclareMathSymbol{\shortminus}{\mathbin}{AMSa}{"39}

\begin{document}
\title{Cavity volume and free energy in many-body systems}
\author[1]{Jamie M. Taylor}
\author[2]{Thomas G. Fai }
\author[3]{Epifanio G. Virga}
\author[4]{Xiaoyu Zheng}
\author[4,5]{Peter Palffy-Muhoray}
\affil[1]{Basque Center for Applied Mathematics (BCAM), Bilbao, Bizkaia, Spain}
\affil[2]{Department of Mathematics and Volen Center for Complex Systems, Brandeis University, Waltham, MA, USA}
\affil[3]{Dipartimento di Matematica, Universit\`a di Pavia, Pavia, Italy}
\affil[4]{Department of Mathematical Sciences, Kent State University, Kent, OH, USA}
\affil[5]{Advanced Materials and Liquid Crystal Institute, Kent State University, Kent, OH, USA}
\date{}
\maketitle
\abstract{\begin{strip}
Within this work we derive and analyse an expression for the free energy of a single-species system in the thermodynamic limit in terms of a generalised cavity volume, that is exact in general, and in principle applicable to systems across their entire range of density, as well as to particles within a general coordinate space. This provides a universal equation of state, and can thus relate the cavity volume to classical results, such as Mayer's cluster expansions. Through this we are able to provide some insight into the connections between cavity volume and free energy density, as well as their consequences. We use examples which permit explicit computations to further probe these results, reclaiming the exact results for a classical Tonks gas and providing a novel derivation of Onsager's free energy for a single species, isotropic system. Given the complexity of the problem we also provide a local lattice ansatz, exact in one dimension, with which we may approximate the cavity volume for hard sphere systems to provide an accurate equation of state in the cases of hard disks and spheres in both dilute regimes as well as beyond the freezing transition. \end{strip}
}

\section{Introduction}

The hard particle system is an attractive toy model of complex systems, as the analysis of the complex energy landscape reduces to purely geometric considerations. Despite the apparent simplicity of such systems, they are capable of demonstrating a rich variety of thermodynamic behaviour such as phase transitions and phase separation \cite{bowick2017mathematics,hansen1990theory,lowen2000fun,santos2020structural}. Results that are both exact and explicit are however generally unavailable, with scarce exceptions such as the exactly solvable one-dimensional Tonks gas \cite{tonks1936complete}. To this end, the predominant tools for analysing hard particle systems are simplified theoretical models and either Monte Carlo or molecular dynamics simulations \cite{alder1957phase,alder1959studies,allen1993hard}. One particularly successful approach is the virial expansion about the vacuum state, which presumes that the equation of state may be written as a power series in the number density, and thus studying the thermodynamic behaviour reduces to finding, either exactly or numerically, the values of the Taylor coefficients \cite{clisby2006ninth, dymond2003virial, wheatley2013calculation}.
Owing to the fact that the virial expansion is a Taylor series about the density zero state, while it may perform incredibly well at low density, at higher densities, or beyond a phase transition, its ability to accurately describe the equation of state deteriorates, and other techniques must be used to describe the system. 

One such method to treat the dense regime is the {\it free volume theory}, which considers the amount of space accessible via continuous movements to an existing particle in the system, \cite{corti1999statistical,hoover1972exact,speedy1980statistical}. The properties of free volume can be related to thermodynamic variables, and have been investigated numerically, typically via the analysis of equilibrium Monte Carlo or molecular dynamics simulations \cite{buehler1951free,hoover1979exact,sastry1998free}. Calculating averages of free volumes is a many-body problem, making explicit closed-form solutions generally unobtainable. One way of vastly simplifying these calculations is the {\it cell theory}, where the system is presumed to be well approximated by a lattice, whose lattice parameters can be derived as a function of the number density, with the consequence that complex multi-particle interactions may be reduced to investigating the local environment of a single particle within the lattice \cite{eyring1937theory,fai2020leaky,hirschfelder1937theory, lennard1937critical,lennard1938critical}. The cell theory is sufficiently simple to provide closed-form results and generally performs well in the dense regime, as dense hard particle systems are lattice-like in two and three dimensions. As this structural assumption is less accurate in dilute systems, however, at lower densities the predictive capability of the model falters.

Within this work we will focus on the cavity volume, which is defined to be the amount of weighted phase space accessible to a new particle within the entire system. In the case of hard particle systems, this simply reduces to the accessible volume, whose complement is the excluded volume, and all quantities of interest become purely geometric in nature. While we shall predominantly focus on hard particle systems, the general framework can equally be applied to soft, long-ranged interactions with sufficient decay at large particle separation. In this case, the weighting of phase space used to calculate the cavity volume corresponds to a Boltzmann factor-type weighting according to the interaction energy. Our goal is to understand how the cavity volume relates to the free energy and equation of state. Geometric properties of the cavity volume for hard spheres in $d$ dimensions have previously been related to thermodynamic variables by Speedy and Reiss, who provided the equation of state
\begin{equation}\label{eqSpeedyReiss}
\frac{P}{\rho kT} = 1 + \frac{\sigma}{2d}\frac{\langle s\rangle}{\langle v\rangle},
\end{equation}
where $\sigma$ is the particle diameter, and $\langle s\rangle, \langle v\rangle$ are the average surface area and volume of a cavity, respectively \cite{speedy1991cavities}. Furthermore, geometric properties of the cavity volume have been experimentally measured in \cite{bowles1994cavities,debenedetti1999statistical, sturgeon1992cavities}.

Within this work we consider two aspects of the problem. In \Cref{secCavityVolume} we present a general study on the cavity volume. By performing thermodynamic integration with respect to the number density, we may relate the cavity volume to the free energy density, equation of state, and virial expansion. We have in mind the case of purely steric interactions between spheres, which will also form the bulk of our later analysis of concrete systems; however, the results of this section are applicable to more general systems, provided the integral defined in \eqref{eqVfDef} exists. This includes cases in which the pair potential depends on internal molecular degrees of freedom such as orientation or conformation, or ``soft" interactions.  

In \Cref{secComputations} we apply these relationships to several model systems, where for definiteness we mostly limit ourselves to familiar cases of hard spheres in one, two and three dimensions. Within this framework we reclaim exact known results for the 1D Tonks gas (\Cref{subsubsecTonks}), a Flory-Huggins-type entropy of mixing which can alternatively be viewed as a linear constitutive assumption on the cavity volume (\Cref{subsubsecLinear}), and a new derivation of the Onsager free energy from an un-correlated, rather than dilute, assumption (\Cref{subsubsecOnsager}). However, in general we will not be so fortunate as to obtain exact results, and thus we are obliged to invoke simplifying assumptions. In light of this, we propose a {\it fluctuating lattice model}, based on an approximation by a local lattice structure that may vary across the domain, which we believe qualitatively captures the significant features of the cavity volume and thus the equation of state, and is exact in one dimension (\Cref{subsecLocalLattice}). Using the fluctuating lattice model, we are able to approximate the cavity volume via a one-dimensional integral that can be computed numerically. This leads to an equation of state which is accurate both in dilute regimes and beyond the freezing transition in the exemplary systems of hard spheres and disks. In particular, this offers advantages over the limitations of the virial expansion and cell theory, which are only valid in dilute and dense regimes, respectively. Comparisons of the obtained equations of state with other models can be found in \Cref{figEOS2DCompare} and \Cref{fig3DEOSComparison} for the two-dimensional and three-dimensional equation of state, respectively.  This is also compared to Monte Carlo estimation of the cavity volume, as outlined in \Cref{subsecMC}. However, in both our theoretical and numerical treatment of the problem we observe a rapid decay of the cavity volume at even moderate densities. This is problematic at the level of evaluating the equation of state, as the logarithm of the cavity volume is the key quantity of interest, meaning that small (absolute) errors in the Monte Carlo scheme can produce catastrophically large errors in the equation of state when the cavity volume is small. Consequently we only consider moderate densities for Monte Carlo experiments.

\section{The cavity volume}\label{secCavityVolume}
\subsection{The free energy density and equation of state}
Let $F_{N,V}$ denote the configurational Helmholtz free energy of a system of $N$ particles with a state space $\Gamma$ of volume $V$, and interacting via a pair potential $U:\Gamma^2\to [0,\infty]$. $\Gamma$ may be simply the positional coordinates of particles, or it may contain internal degrees of freedom such as orientation. Explicitly, we may write the configurational partition function of $N$ particles in the volume $V$, denoted $\mathcal{Z}_{N,V}$, as
\begin{equation}
\mathcal{Z}_{N,V}=\frac{1}{N!}\int_{\Gamma^N}e^{-\frac{1}{kT}\sum\limits_{1\leq i<j\leq N} U(q_i,q_j)}\,d^N\bar{q}_N,
\end{equation}  
where $\bar{q}_N=(q_1,q_2,...,q_N)\in \Gamma^N$. We then have the configurational Helmholtz free energy given as 
\begin{equation}
F_{N,V}=-kT\ln \mathcal{Z}_{N,V}.
\end{equation} 
with $k$ the Boltzmann constant and $T>0$ the absolute temperature. It is known, and a cornerstone of the Widom insertion method \cite{widom1963some} (see also \cite{jackson1964potential}), that 
 \begin{equation}\label{eqInsertion}\begin{split}
 &\mathcal{Z}_{N,V}\\
 &=\frac{1}{N}\mathcal{Z}_{N-1,V}\left\langle \int_\Gamma e^{-\frac{1}{kT}\sum\limits_{i=1}^{N-1} U(q,q_i)}\,dq\right\rangle_{P_{N-1,V}},
 \end{split}
 \end{equation}
 where the average is taken over particles $\bar{q}_{N-1}$ according to the Gibbs distribution $P_{N-1,V}$. In this work, we express Widom's relationship using a different but equivalent set of variables. For a system of $N-1$ particles in a volume $V$ corresponding to the interaction $U$, $P_{N-1,V}$ is given as
\begin{equation}\begin{split}
&P_{N-1,V}(\bar{q}_{N-1})\\
=&\frac{1}{\mathcal{Z}_{N-1,V}}e^{-\frac{1}{kT}\sum\limits_{1\leq i<j\leq N-1}U(q_i,q_j)}.
\end{split}
\end{equation}
Following \eqref{eqInsertion}, we have the recurrence relation
\begin{equation}\begin{split}
&F_{N,V}-F_{N-1,V}\\
&=kT\ln N - kT\ln\left\langle \int_\Gamma e^{-\frac{1}{kT}\sum\limits_{i=1}^{N-1} U(q,q_i)}\,dq\right\rangle_{P_{N-1,V}} .
\end{split}
\end{equation} For brevity, we will denote 
\begin{equation}\label{eqVfDef}
V_f(N,V)=\left\langle  \frac{1}{V}\int_\Gamma e^{-\frac{1}{kT}\sum\limits_{i=1}^{N-1} U(q,q_i)}\,dq\right\rangle_{P_{N-1,V}},
\end{equation}
which we interpret to be the generalised cavity volume fraction, with a possible weighting corresponding to soft interactions. In the case of purely steric interactions, where $U(q_i,q_j)=+\infty$ if particles with coordinates $q_i,q_j$ intersect, and $0$ otherwise, this is precisely the volume fraction available to a new particle, randomly sampled according to the integration measure on $\Gamma$.

This allows us to calculate the total free energy density $\mf_{N,V}=\frac{F_{N,V}}{V}$ by an incremental scheme, as 
\begin{equation}\label{eqIncrementSum}
\begin{split}
&\mf_{N,V}\\
=&\frac{1}{V}F_{0,V}+\sum\limits_{n=1}^N \frac{1}{V}(F_{n,V}-F_{n-1,V})\\
=&\mf_{0,V}+kT\sum\limits_{n=1}^N \frac{1}{V}\left(\ln n-\ln V V_f(n-1,V)\right).
\end{split}
\end{equation}
We presume that $V_f$ is well defined in the thermodynamic limit, that is,  there exists some function $\mathcal{V}_f$ so that for $N,V$ large, $V_f(N,V)\approx \mathcal{V}_f\left(\frac{N}{V}\right)$. If $\ln\mathcal{V}_f$ is Riemann integrable, we may approximate the sum \eqref{eqIncrementSum} as a Riemann sum with step size $\Delta x=\frac{1}{V}$, $x_n=n\Delta x$. Furthermore taking the vacuum energy $F_{0,V}=0$, without loss of generality, and replacing $V_f$ with $\mathcal{V}_f$, this reduces to
\begin{equation}\label{eqFreeEnergy}
\begin{split}
\frac{\mf_{N,V}}{kT}\approx &\sum\limits_{n=1}^N \frac{1}{V}\left(\ln \frac{n}{V}-\ln  \mathcal{V}_f\left(\frac{n}{V}\right)\right)\\
\approx &\int_0^{\frac{N}{V}}\ln(x)-\ln \mathcal{V}_f(x)\,dx\\
=& \rho \ln \rho - \rho - \int_0^\rho \ln \mathcal{V}_f(x)\,dx,
\end{split}
\end{equation}
where $\frac{N}{V}=\rho$ denotes the number density. Following this, we define the free energy density $\mathcal{F}$ at number density $\rho$ as 

\begin{equation}\label{eqDefEnergy}
\mathcal{F}(\rho):=kT\left(\rho \ln \rho - \rho - \int_0^\rho \ln \mathcal{V}_f(x)\,dx\right).
\end{equation}

We expect the function $\mathcal{V}_f$ to satisfy $\mathcal{V}_f(0)=1$ and $\mathcal{V}_f(\rho^*)=0$ where $\rho^*$ is the highest possible number densit. We do not rule out the possibility of non-monotonicty of $\mathcal{V}_f$ between these regimes. In the case of purely steric interactions, where $U$ attains only the values zero (no penetration of particles) or $\infty$ (penetration of particles), $\mathcal{V}_f$ only depends on the number density and particle shape. However in the case of soft interactions, in which $U$ attains values other than $0$ and $+\infty$, $\mathcal{V}_f$ will also be dependent on the temperature, which is immediate from \eqref{eqVfDef}.

Equation \eqref{eqDefEnergy} shows the significance of the cavity volume fraction in calculating the energy, and we can provide a universal equation of state for such a system as 
\begin{equation}\label{eqEOSUniversal}\begin{split}
\frac{P}{kT}= &\frac{1}{kT}\left(-\mf(\rho)+\rho\frac{\partial \mf}{\partial \rho}(\rho)\right)\\
=&\rho-\rho\ln \mathcal{V}_f(\rho)+\int_0^\rho \ln \mathcal{V}_f(x)\,dx. 
\end{split}
\end{equation}
By integration by parts if $\mathcal{V}_f$ is differentiable, we may also write that 
\begin{equation}\label{eqEOSfunky}
\frac{P}{\rho kT}=1-\frac{1}{\rho}\int_0^\rho\frac{x\mathcal{V}_f'(x)}{\mathcal{V}_f(x)}\,dx.
\end{equation}
This formula is particularly interesting as it has a superficial similarity to the Speedy and Reiss formula \eqref{eqSpeedyReiss} for hard spheres. In dilute systems when the excluded volume is made entirely of disjoint exclusion spheres, $\mathcal{V}_f(\rho)=1-\rho\sigma^d\omega_d$, where $\omega_d$ is the volume of a ball of radius $1$ in $\mathbb{R}^d$. Then the surface area of this excluded volume is $d\rho\sigma^{d-1}\omega_d$, so that $\frac{\sigma}{2d}\langle s\rangle =\frac{1}{2}\rho\sigma^d\omega_d=-\frac{\rho}{2} \mathcal{V}_f'(\rho)$. Roughly speaking, this suggests that \eqref{eqEOSfunky} may be a kind of integral formulation of the Speedy formula.

It is possible to invert the equation \eqref{eqEOSfunky} to provide the cavity volume as a function of pressure, if the latter is known. It is immediate that by rearranging and taking a derivative, we have the relationship
\begin{equation}
1-\frac{\partial}{\partial\rho}\frac{P(\rho)}{kT}=\frac{\rho\mathcal{V}_f'(\rho)}{\mathcal{V}_f(\rho)}=\rho\frac{\partial}{\partial\rho}\ln\mathcal{V}_f(\rho).
\end{equation}
By further rearrangement and integration, we have that 
\begin{equation}\label{eqInvertEOS}
\mathcal{V}_f(\rho)=\exp\left(\int_0^\rho \frac{1}{x}\left(1-\frac{1}{kT}\frac{\partial P}{\partial x}\right)\,dx\right).
\end{equation}

The cavity volume fraction may also be expressed in terms of the compressibility factor, 
\begin{equation}
Z:=\frac{P}{\rho kT},
\end{equation}
which is typically defined in terms of the packing fraction $\eta$, defined by
\begin{equation}
\eta:=\rho v_0,
\end{equation}
with $v_0$ the single particle volume, giving 
\begin{equation}\label{eqInvertEOSPacking}
\mathcal{V}_f(\eta)=\exp\left(\int_0^\eta \frac{1}{x}\left(1-\frac{\partial}{\partial x}(xZ(x))\right)\,dx\right).
\end{equation}
For example, by inserting into \eqref{eqInvertEOSPacking} the equation of state for Carnahan-Starling (see, e.g., \cite[Page 76]{santos2016concise}), we obtain 
\begin{equation}
\mathcal{V}_f(\eta)=\exp\left(\frac{\eta(3\eta^2-9\eta+8)}{(1-\eta)^3}\right)
\end{equation}

Furthermore, we can infer from \eqref{eqDefEnergy} that if $\mathcal{V}_f$ is $k$-times differentiable and non-zero, $\mf$ is consequently $k+1$-times differentiable. More so, if $\mathcal{V}_f$ is continuous and monotonically decreasing, then $\frac{\partial\mf}{\partial \rho}=kT\ln\rho-1-kT\ln \mathcal{V}_f(\rho)$ is necessarily continuous and monotonically increasing, implying that $\mf$ is convex, and thus there can be no phase separation.

This approach is essentially a form of thermodynamic integration \cite[Section 7.1]{frenkel2001understanding}, where the free energy difference between two states is calculated by integrating a tractable derivative of the energy between the states. In our case, we are calculating the energy difference between the vacuum and a given state by integrating the chemical potential as a function of number density in the thermodynamic limit. A standard application of this technique is to consider two systems with different interactions, parametrise a path between the two types of interaction, and integrate the derivative of the free energy across this path to evaluate the energy difference of the two systems. Typically thermodynamic integration is performed numerically, with the derivative of the free energy estimated via simulation methods. The advantage of such a scheme is that the quadrature scheme implicitly averages the errors of constituent simulations when evaluating the energy difference, which consequently makes the final results typically more accurate than the individual simulations used to obtain them.
\subsection{Virial expansion and cavity volume}
The virial equation of state for a system is 
\begin{equation}
\frac{1}{kT}P=\sum\limits_{n=1}^\infty \textup{B}_n\rho^n,
\end{equation}
where the constants $\textup{B}_n$ are known as the virial coefficients. There is a rich history of attempts to compute and approximate the virial coefficents \cite{dymond2003virial,kamerlingh1901expression}. In this discussion we will relate the virial coefficients to the cavity volume in an exact and general way, expressing the virial coefficients in terms of an analogous series expansion for the cavity volume fraction.

The coefficients $\textup{B}_n$ have the same units of $\rho^{1-n}$ and $v_0^{n-1}$, thus we introduce the {\it reduced} virial coefficients $\text{b}_n$, defined as $\text{b}_n=v_0^{1-n}\textup{B}_n$. The reduced virial coefficients $\text{b}_n$ are thus dimensionless constants that satisfy 
\begin{equation}
\frac{P}{\rho kT}=Z=\sum\limits_{n=0}^\infty \text{b}_{n+1}\eta^{n},
\end{equation}
with $\eta=\rho v_0$ the packing fraction as before. 

For the following it is more convenient to work with the excluded volume than the cavity volume, with the excluded volume fraction $\bar{\mathcal{V}}_f$ defined as 
\begin{equation}
1-\bar{\mathcal{V}}_f(\rho)=\mathcal{V}_f(\rho)
\end{equation} Assume that the excluded volume is analytic at the dilute regime, so that it may be written as 
\begin{equation}
\bar{\mathcal{V}}_f(\rho)=\sum\limits_{n=1}^\infty \frac{v_n\rho^n}{n!}
\end{equation}
for $\rho$ sufficiently close to $0$ and coefficients $v_n$. We note that as $\mathcal{V}_f(0)=1$, $\bar{\mathcal{V}}_f(0)=0$, hence the sum is over $n\geq 1$. 

In the following we will use $B_{n,k}$ to denote the partial exponential Bell polynomials. For an introduction and discussion on Bell polynomials the reader is directed to \cite[Section 3.3]{comtet2012advanced}. These may be defined in numerous ways, one of which is via a generating function type equality, so that they satisfy the relationship 
\begin{equation}
\frac{1}{k!}\left(\sum\limits_{j=1}^\infty x_j\frac{t^j}{j!}\right)^k=\sum\limits_{n=k}^\infty B_{n,k}(x_1,x_2,...,x_{n-k+1})\frac{t^n}{n!}
\end{equation} 
for real $t$, integer $k$ and sequences $(x_j)_{j=1}^\infty$ so that the sums are absolutely convergent. Immediately following the definition, we see that if $y_j=\frac{x_j}{\alpha^j}$, then $y_j(\alpha t)^j=x_jt^j$. This implies a homogeneity condition 
\begin{equation}\label{eqHomoBell}
\alpha^n B_{n,k}(y_1,y_2,...,y_{n-k+1})=B_{n,k}(x_1,x_2,...x_{n-k+1}). 
\end{equation}
For notational brevity, we consider the excess free energy $\bar{\mathcal{F}}$, defined as 
\begin{equation}
\bar{\mathcal{F}}(\rho)=\mathcal{F}(\rho)-kT(\rho\ln\rho-\rho).
\end{equation}
This is simply the non-ideal part of the free energy. $\bar{\mathcal{F}}$ may then be written as
\begin{equation}
\begin{split}
& \frac{1}{kT}\bar{\mathcal{F}}= \\
& -\int_0^\rho \ln\big(1-\bar{\mathcal{V}}_f(t)\big)\,dt\\
=& \int_0^\rho \sum\limits_{k=1}^\infty \frac{\bar{\mathcal{V}}_f(t)^k}{k}\,dt\\
=&\int_0^\rho \sum\limits_{k=1}^\infty (k-1)! \frac{1}{k!}\left(\sum\limits_{n=1}^\infty \frac{v_nt^n}{n!}\right)^k\,dt\\
=& \int_0^\rho \sum\limits_{k=1}^\infty (k-1)!\sum\limits_{n=k}^\infty B_{n,k}(v_1,...,v_{n-k+1})\frac{t^n}{n!}\,dt\\
=& \int_0^\rho  \sum\limits_{n=1}^\infty\left(\sum\limits_{k=1}^n(k-1)! B_{n,k}(v_1,...,v_{n-k+1})\right)\frac{t^n}{n!}\,dt\\
=&\sum\limits_{n=1}^\infty\left(\sum\limits_{k=1}^n (k-1)!B_{n,k}(v_1,...,v_{n-k+1})\right)\frac{\rho^{n+1}}{(n+1)!}.
\end{split}
\end{equation}
This gives the virial equation of state as 
\begin{equation}
\begin{split}
 &\frac{P}{\rho kT}-1= \\
&-\frac{1}{\rho kT}\bar{\mathcal{F}}+\frac{1}{kT}\frac{\partial\bar{\mathcal{F}}}{\partial \rho}\\
=&\sum\limits_{n=1}^\infty\left(\sum\limits_{k=1}^n (k\shortminus 1)!B_{n,k}(v_1,...,v_{n\shortminus k+1})\right)\frac{n\rho^{n}}{(n+1)!},
\end{split}
\end{equation}

which in turn gives the virial coefficients as 
\begin{equation}\label{eqVirialVolume}
\textup{B}_{n}=\frac{(n\shortminus 1)}{n!}\left(\sum\limits_{k=1}^{n-1} (k\shortminus 1)!B_{n\shortminus 1,k}(v_1,...,v_{n\shortminus k})\right).
\end{equation}
We note that this expression is valid for $n\geq 2$, as the ideal gas term corresponding to $n=1$ is not considered.

Dividing through by $v_0^{n-1}$ to give the reduced virial coefficients, using the homogeneity condition \eqref{eqHomoBell} gives 
\begin{equation}
\text{ b}_n=\frac{n\shortminus 1}{n!}\left(\sum\limits_{k=1}^{n\shortminus 1} (k\shortminus 1)!B_{n\shortminus 1,k}(u_1,...,u_{n\shortminus k})\right).
\end{equation}
where $u_j = v_0^{-j}v_j$, so that for $\eta=\rho v_0$, 
\begin{equation}
\bar{\mathcal{V}}_f(\rho)=\sum\limits_{i=1}^\infty\frac{v_i\rho^i}{i!}=\sum\limits_{i=1}^\infty \frac{u_i\eta^i}{i!}.
\end{equation}

In principle, the coefficients $v_k$ can be computed numerically. This is because for $V$ sufficiently large, we know that $\mathcal{V}_f\left(\frac{N}{V}\right)\approx V_{f}\left(N,V\right)$, and $V_{f}$ is a quantity defined for finite systems. Using this, we can approximate derivatives at $0$ by finite differences with step sizes $h=\frac{1}{V}$, giving 
\begin{equation}\label{eqFiniteDifference}
\begin{split}
v_k=&\left.\frac{d^k}{d\rho^k}\bar{\mathcal{V}}_f(\rho)\right|_{\rho=0}\\
\approx & V^k\sum\limits_{i=1}^k \binom{k}{i} \bar{\mathcal{V}}_f\left(\frac{i}{V}\right)(-1)^{k-i}\\
\approx & V^{k}\sum\limits_{i=1}^k \binom{k}{i} \left(1-V_f\left(i,V\right)\right)(-1)^{k-i}\\
\end{split}
\end{equation}

The first of the approximations should be exact as $V\to \infty$ if $\mathcal{V}_f$ is sufficiently smooth, and the second if $V_f(i,V)$ is sufficiently well approximated by $\mathcal{V}_f\left(\frac{i}{V}\right)$. This implies that to find the virial coefficient $\text{B}_n$, in principle it suffices to have a very good estimate of $V_f(i,V)$ for $i=1,...,n-1$, which amounts to estimating how much space is excluded in a system of $n-1$ particles. 

In terms of the dimensionless coefficients $u_k$, the first few reduced virial coefficients are thus given by 
\begin{equation}
\begin{split}
{\text b}_2=&\frac{u_1}{2},\\
{\text b}_3=& \frac{1}{3}\left(u_1^2+u_2\right),\\
{\text b}_4=& \frac{1}{8}\left(2u_1^3+3u_1u_2+u_3\right),\\
{\text b}_5=&\frac{1}{30}\left(6u_1^4+12u_1^2u_2+3u_2^2+4u_1u_3+u_4\right).
\end{split}
\end{equation}

Using $V_i=V_f(i,V)$ for brevity, we can thus give the virial coefficients in terms of $i$-particle excluded volumes as 
\begin{equation}
\begin{split}
{\text b}_2=&\frac{V}{2v_0}(1-V_1),\\
{\text b}_3=& \frac{V^2}{3v_0^2}\left(V_1^2-2V_2\right),\\
{\text b}_4=& \frac{V^3}{8v_0^3}\left(-2V_1^3+3V_1V_2-V_3\right),\\
{\text b}_5=& \frac{V^4}{30v_0^4}\left(6V_1^4-12V_1^2V_2+3V_2^2+4V_1V_3-V_4\right).
\end{split}
\end{equation}
There will be three sources of error in using such representations to numerically obtain virial coefficients, which are finite size effects in a sampled system, sampling and integration error in evaluating $V(i,V)$, and error in the approximations of the derivatives.

\subsection{Comparison with the free volume theory}\label{subsecFreeVolume}

The theory presented within this work concerns cavity volume, which is related, although distinct from, the well studied theory of {\it free volume }\cite{buehler1951free,eyring1937theory,fai2020leaky,lennard1937critical, lennard1938critical}. Given a configuration of hard particles, the free volume of a particular particle is the volume that it may access by continuous motion while holding all other particles in place, without particle interpenetration. We emphasise that the free volume, in contrast to the cavity volume, is not a globally defined property of the system; rather, it is a region of space defined {\it with respect to} a probe particle within the system. Regions of free volume with respect to different probe particles may be disjoint, or have a non-empty intersection. In a dilute regime, where typically particles are typically far apart, the particle can access almost any space outside of the exclusion spheres of other particles, which are disjoint. This implies that the free volume is an extensive quantity roughly equal to $V-N\vo$, where $\vo$ is the pairwise excluded volume of an average particle. In dense systems however, particles are expected to be caged by their neighbours, in which case the free volume is an intensive quantity, and the size of the free volume should be comparable to the size of the particle itself. The qualitative change of the free volume from an extensive to intensive quantity is associated with the percolation transition. The advantage of the free volume theory is that it is relatively straightforward to analyse in dense systems, and the geometric properties of the free volume can be related to various thermodynamic quantities. As noted by Sastry {\it et al.} \cite{sastry1998free}, however, the free volume is linked with the cavity volume in an intimate way. We replicate the key points from their discussion here for comparison.

We define a {\it connected cavity} to be a connected component of the cavity volume, and we will denote the volume of a connected cavity by $u$. Furthermore, given a particular particle we denote its free volume as $v$. Then in a system of $N+1$ particles, the free volume of the $N+1$-th particle corresponds to a connected cavity in a system of $N$ particles, and the $N+1$-th particle will be added to any particular connected cavity with a probability proportional to its size, $u$. This argument can be used to show that the average free volume and average cavity volume can be related by 
\begin{equation}
\langle v^{-1}\rangle^{-1}=\langle u\rangle.
\end{equation}
That is, the arithmetic mean of the connected cavity size is equal to the harmonic mean of the free volume. If we have $N_c$ connected cavities in a typical system, we may then relate the total cavity volume with the harmonic mean of the free volume by the relationship
\begin{equation}
V_f(N,V)=1-\frac{N_c}{V}\langle u\rangle =1-\frac{N_c}{V\langle v^{-1}\rangle}. 
\end{equation}
As $V$ is extensive and $V_f$ is intensive, this implies that $\frac{N_c}{\langle v^{-1}\rangle}$ is also an extensive quantity, and thus at the percolation transition, when $v$ undergoes a qualitative change from an extensive to intensive quantity, we must have that $N_c$ does the reverse, going from an intensive quantity (equal to $1$ in the dilute limit for hard sphere systems in dimension greater than 1) to an extensive quantity. This illustrates that while there is a link between the two theories, there is a further quantity, $N_c$, which obscures the relationship between the two and remains elusive. Furthermore, as both $N_c$ and $ v$ undergo an extensive/intensive exchange as the concentration increases and decreases (respectively), any attempt to define them in the thermodynamic limit for all concentrations would prove difficult. This particular problem is not present in the cavity volume approach, as the cavity volume fraction is always an intensive variable and can thus be defined without complication in the thermodynamic limit. In particular, this allows us to use the cavity volume approach across the entire range of number densities without the need to impose different ad-hoc definitions in different regimes.

\subsection{$\lambda$-function}\label{subsecLambda}
In \cite{nascimento2017density} a constitutive equation for the excluded volume fraction was proposed, of the form
\begin{equation}
1-\mathcal{V}_f(\rho)=\lambda(\rho)\vo\rho,
\end{equation}
where $\vo$ is the pairwise excluded volume given by 
\begin{equation}\label{eqExcludedVolume}
\vo=\int_{\mathbb{R}^d}1-\exp\left(-\frac{1}{kT}U(q_1,q)\right)\,dq,
\end{equation}
which is implicitly assumed to be independent of $q_1$ on symmetry grounds. The introduction of the function $\lambda$ was mostly ad-hoc, and aimed to act as an effective correction for the over-counting of pairwise excluded volume. Within this section we aim to revisit $\lambda$ and its interpretation. 

Consider the case of a pair potential $U$, taken to be non-negative without loss of generality. Recall we may write the (weighted) cavity volume fraction in a single species system as 
\begin{equation}
V_f(N,V)=\frac{1}{V}\left\langle \int_\Gamma e^{-\frac{1}{kT}\sum\limits_{i=1}^{N-1} U_{i,N}}\,dq\right\rangle_{P_{N-1}}.
\end{equation}
We consider arbitrary particles with generalised coordinates $q\in \Gamma$. Following the traditions of cluster expansions \cite{mayer1937statistical} define the new variable $f_i$, analogous to a Mayer function, as 
\begin{equation}
f_i=1-\exp\left(-\frac{1}{kT}U_{i,N}\right),
\end{equation}
where for brevity we denote $U(q_i,q_N)=U_{i,N}$. Then we may re-write the cavity volume fraction as 
\begin{equation}\begin{split}
&1-V_f(N,V)\\
=&1-\frac{1}{V}\int_\Gamma\left\langle\prod\limits_{i=1}^{N-1} (1-f_i)\right\rangle\,dq\\
=&\frac{1}{V} \int_\Gamma\left\langle {\textstyle\sum\limits_{i_1}} f_{i_1}-\sum\limits_{i_1,i_2} f_{i_1}f_{i_2}\right.\\
&\hspace{1cm}\left.+{\textstyle\sum\limits_{i_1,i_2,i_3}} f_{i_1}f_{i_2}f_{i_3}+...\right\rangle\,dq\\
=&  {\textstyle\sum\limits_{i_1}}\frac{1}{V}\int_\Gamma \left\langle f_{i_1}\left( 1-{\textstyle\sum\limits_{i_2}} f_{i_2}+{\textstyle\sum\limits_{i_2,i_3} }f_{i_2}f_{i_3}+...\right)\right\rangle\,dq
\end{split}
\end{equation}
The sums are taken over $i_j=1,...,N-1$, and exclude cases where $i_{j}=i_{k}$ for $j\neq k$, and averages $\langle\cdot\rangle$ are with respect to $P_{N-1}$. As $f_{i_1}P_{N-1}$ is non-negative, if the series in the brackets is bounded above and below we may find a dimensionless value $\lambda_{i_1}>0$ so that 
\begin{equation}\begin{split}
&\int_\Gamma \left\langle f_{i_1}\left( 1-{\textstyle\sum\limits_{i_2=1}^{N-1}} f_{i_2}+{\textstyle\sum\limits_{i_2,i_3=1}^{N-1}} f_{i_2}f_{i_3}-...\right)\right\rangle\,dq\\
=&\lambda_{i_1}\int_\Gamma \left\langle f_{i_1}\right\rangle_{P_{N-1}}\,dq
\end{split}
\end{equation}

If the energy $U$ is continuous, we may apply the mean value theorem for products of integrals to conclude that there exists some $(q_1,q_2,...,q_{N-1})\in\Gamma^{N-1}$, $q\in \Gamma$ such that the scalar $\lambda_i$ precisely satisfies 
\begin{equation}
\begin{split}
\lambda_{i_1}&=\left( 1-{\textstyle\sum\limits_{i_2=1}^{N-1}} f_{i_2}+{\textstyle\sum\limits_{i_2,i_3=1}^{N-1}} f_{i_2}f_{i_3}-...\right),\\
f_{j}=&1-\exp\left(-\frac{1}{kT}U(q_{j},q)\right).
\end{split}
\end{equation}
Following this we may interpret $\lambda_{i_1}$ as a ``typical" re-counting factor for overlaps, and in the discontinuous case we may still view it as a kind of effective recount. By symmetry, $\lambda_{i_1}$ may only be a function of number density. That is, $\lambda_{i_1}=\lambda(\rho)$. Hence we may write that 
\begin{equation}\begin{split}
&V_f(N,V)\\
=&1-\lambda(\rho)\sum\limits_{i_1=1}^{N-1}\frac{1}{V}\int_\Gamma \langle f_{i_1}\rangle_{P_{N-1}}\,dq\\
=&1-\lambda(\rho)\frac{N-1}{V}\vo.\end{split}
\end{equation}
This then gives the cavity volume fraction in the thermodynamic limit as 
\begin{equation}\label{eqLambdaExpr}
\mathcal{V}_f(\rho)=1-\lambda(\rho)\rho \vo.
\end{equation}
Of course $\lambda$ can be readily claimed from a known expression for $\mathcal{V}_f(\rho)$ provided $\vo,\rho$ are known, but we claim its interpretation as an effective correction for overlapping exclusion zones makes it a worthwhile quantity to consider in itself. 

In the case of hard spheres in $n$-dimensions, $\vo=2^nv_0$, so the fact that $\mathcal{V}_f\in[0,1]$ implies that $\lambda(\rho) \in [2^{-n},1]$, where we expect $\lambda(0)=1$.

\section{Applications and results}\label{secComputations}

We now turn towards the application of our previous general results to concrete systems. Our main novel contribution will be our fluctuating lattice model, which is defined and analysed in \Cref{subsecLocalLattice}. Before addressing this however, we aim to use the following subsection to illustrate uses of the framework developed throughout \Cref{secCavityVolume} within more familiar contexts to demonstrate its efficacy in obtaining several classical results. Explicitly, we may calculate exactly the cavity volume to rederive the equation of state for the Tonks gas in one dimension (\Cref{subsubsecTonks}), use a linear constitutive equation on the cavity volume to obtain the Flory-Huggins entropy of mixing (\Cref{subsubsecLinear}), and take an ansantz of an uncorrelated system to obtain the virial expansion to second order (\Cref{subsubsecOnsager}).

\subsubsection{Tonks Gas}\label{subsubsecTonks}
A one-dimensional system of hard rods on a line is an exactly solvable system \cite{tonks1936complete}. Here we verify that classical results may be re-obtained via calculation of the cavity volume and \eqref{eqDefEnergy}. More so, given that many exact computations are possible for this one-dimensional system, we can use this as a test case to understand in more depth the nature of cavity volume. Speedy and Reiss \cite{speedy1991cavities} performed similar calculations for evaluating a distinct quantity, the free volume. As the calculations we perform here are near-equivalent, we direct the reader to their work for the details and provide only the key steps. We first note a system of $N$ hard rods of length $\sigma$ on a periodic line (or equivalently, a circle) of length $V$ is equivalent to $N$ non-interacting points $(x_i)_{i=1}^N$ on a periodic line of length $V-N\sigma$. Given some probe point $x$, the periodic distances from $x$ are then distributed uniformly on $\frac{V-N\sigma}{2}$. Thus the probability that all particles are at a distance at least $r>0$ from $x$ is given by
\begin{equation}
\left(\frac{V-N\sigma-2r}{V-N\sigma}\right)^N.
\end{equation}
In particular, taking $r=\frac{\sigma}{2}$, we see that the probability of being able to insert a new particle of length $\sigma$ on the line of length $V-N\sigma$ is 
\begin{equation}\label{eqTonksProb}
\left(\frac{V-N\sigma-\sigma}{V-N\sigma}\right)^N=\left(1-\frac{\sigma}{V-N\sigma}\right)^N.
\end{equation}
This then gives the cavity volume fraction on a line of length $V$ as 
\begin{equation}
V_f(N,V)=\left(1-\frac{\sigma}{V-N\sigma}\right)^N\left(\frac{V-N\sigma}{V}\right),
\end{equation}
where the latter bracket corresponds to the probability of a probe point not lying within an existing particle. Introducing the number density $\rho=\frac{N}{V}$, we see this has a readily computable limit as $N,V\to \infty$ with $\frac{N}{V}=\rho$ given by
\begin{equation}
\begin{split}
V_f(N,V)=&\left(1-\frac{\sigma}{V-N\sigma}\right)^N\left(\frac{V-N\sigma}{V}\right)\\
=& \left(1-\frac{\rho\sigma}{N(1-\rho\sigma)}\right)^N\left(1-\rho\sigma\right)\\
\to & \exp\left(-\frac{\rho\sigma}{1-\rho\sigma}\right)\left(1-\rho\sigma\right). 
\end{split}
\end{equation}
This expression decays to zero faster than any polynomial as $\rho\to \frac{1}{\sigma}$, in fact it is a classical example of a function where every (left-)derivative vanishes at $\rho=\frac{1}{\sigma}$, but the function is non-zero for $\rho<\frac{1}{\sigma}$. In particular, a truncated series expansion of the cavity volume at $\rho=\frac{1}{\sigma}$ would only produce $\mathcal{V}_f(\rho)=0$, giving no information.

We may then calculate 
\begin{equation}
\begin{split}
&\int_0^\rho \ln\mathcal{V}_f(x)\,dx\\
=&\int_0^\rho \ln\left(\exp\left(-\frac{x\sigma}{1-x\sigma}\right)(1-x\sigma)\right)\,dx\\
=& \int_0^\rho \frac{-x\sigma}{1-x\sigma}+\ln(1-x\sigma)\,dx\\
=& \rho\ln(1-\rho\sigma). 
\end{split}
\end{equation}
By \eqref{eqDefEnergy}, we obtain a logarithmic singularity in the free energy density,
\begin{equation}
\frac{1}{kT}\mf=\rho\ln\rho-\rho-\rho\ln(1-\rho\sigma), 
\end{equation}
and the classical equation of state for a Tonks gas, 
\begin{equation}
P=-\mf+\rho\frac{\partial\mf}{\partial\rho}=\frac{kT\rho}{1-\rho\sigma}. 
\end{equation}
By means of the free energy we can observe that there are no phase transitions as the energy is analytic in the number density for $0<\rho<\frac{1}{\sigma}$, and we can rule out phase separation as the energy is strictly convex, as verified from its second derivative 
\begin{equation}
\frac{1}{kT}\frac{\partial^2}{\partial\rho^2}\mf=\frac{1}{\rho(1-\rho\sigma)^2}>0.
\end{equation}

\begin{figure}[H]
\begin{center}
\includegraphics[width=0.45\textwidth]{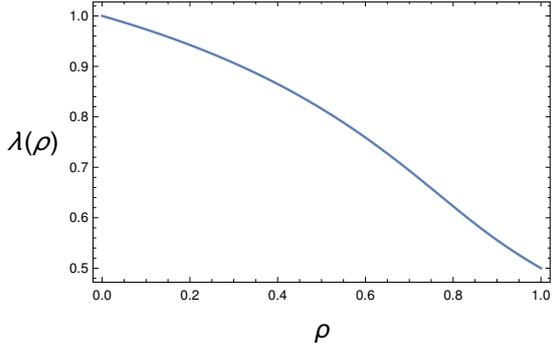}
\caption{$\lambda(\rho)$ versus $\rho$ for $\sigma=1$.}\label{figLambdaTonks}
\end{center}
\end{figure}

We can reclaim $\lambda$, as per \Cref{subsecLambda}, from the cavity volume as 
\begin{equation}
\begin{split}
\lambda(\rho)=& \frac{1-\exp\left(-\frac{\rho\sigma}{1-\rho\sigma}\right)(1-\rho\sigma)}{2\rho\sigma},
\end{split}
\end{equation}
and for which we have included a plot in Figure \ref{figLambdaTonks}. We observe a monotonic, smooth decay from $\lambda(0)=1$ to $\lambda\left(\frac{1}{\sigma}\right)=\frac{1}{2}$. 

We can consider the total number of cavities, $N_c$, as defined in \Cref{subsecFreeVolume}. It should be noted that $N_c$ is of a very different nature in one dimension, as a single particle disconnects space. Because of this, $N_c$ has limiting behaviour $N_c\to N$ in the dilute regime, and is always extensive. This is in contrast to the behaviour of spheres in $d>1$ dimensions, which would have $N_c\to 1$. Using the same argument as used to obtain \eqref{eqTonksProb}, we see that a cavity exists between a particle and its neighbour if and only if they are separated by a distance of at least $2r=\sigma$, which has probability
\begin{equation}
\left(\frac{V-(N+1)\sigma}{V-N\sigma}\right)^N.
\end{equation}
The number of cavities {\it per volume} is then simply this quantity multiplied by $\rho=\frac{N}{V}$, giving
\begin{equation}\begin{split}
&\frac{1}{V}N_c\\
=&\frac{N}{V}\left(\frac{V-(N+1)\sigma}{V-N\sigma}\right)^N \\
\to &\rho\exp\left(-\frac{\rho\sigma}{1-\rho\sigma}\right).
\end{split}
\end{equation}
Furthermore, we can then infer the average cavity size and harmonic mean of the free volume as 
\begin{equation}
\langle u\rangle =\langle v^{-1}\rangle^{-1}=\frac{V\mathcal{V}_f}{N_c}=\frac{1-\rho\sigma}{\rho}
\end{equation}

In the dense limit, as $\rho\nearrow\rho^*=\frac{1}{\sigma}$, we thus see that the leading contribution to the decay of the cavity volume is not from the {\it size} of individual cavities, but their {\it rarity}, as $\langle u\rangle$ converges linearly in $\rho^*-\rho$ to zero, while $N_c$ decays faster than any polynomial. 

\subsubsection{Linear Depletion and hole/particle equivalence}\label{subsubsecLinear}
Outside of some toy systems, we generally cannot expect to evaluate the cavity volume exactly, so we will need to invoke an ansatz or constitutive equation on the cavity volume. The simplest such ansatz we may take is that the cavity volume is an affine function of number density. To agree with exact results in the dilute limit, we take the constitutive equation
\begin{equation}\label{eqLinearDepletion}
\mathcal{V}_f(\rho)=1-\rho\vo,
\end{equation}
where $\vo$ is the two-particle generalised excluded volume, as in \eqref{eqExcludedVolume}. This constitutive equation can be viewed in one of two manners; the first being that it is a linear approximation to the excluded volume of a general system, the second is that this is the exact behaviour of a system of finitely many discrete ``bins" that can each hold at most one particle. In terms of the previously defined $\lambda$-function, this corresponds to the constitutive equation $\lambda(\rho)=1$. In the case of the constitutive equation \eqref{eqLinearDepletion}, the free energy can be evaluated explicitly, as 
\begin{equation}
\begin{split}
\int_0^\rho \ln(\mathcal{V}_f(x))\,dx=& -\rho-\frac{1-\rho \vo}{\vo}\ln(1-\rho \vo).
\end{split}
\end{equation}
Introducing a rescaled number density $\xi=\rho \vo$, we then have the free energy via \eqref{eqDefEnergy} as 
\begin{equation}\label{eqLinearEnergy}\begin{split}
&\frac{1}{kT}\mf\\
=& \rho\ln\rho+\frac{1-\rho \vo}{\vo}\ln(1-\rho \vo)\\
=& \frac{1}{\vo}\left(\xi\ln\xi+(1-\xi)\ln(1-\xi)\right)-\xi\frac{\ln\vo}{\vo}.
\end{split}
\end{equation}
The system is saturated as $\rho\to\frac{1}{\vo}^-$, and the energy is bounded in this limit, with 
$$\lim\limits_{\rho\to\frac{1}{\vo}^-}\mf=-kT\frac{\ln \vo}{\vo}.$$ 
This free energy gives the equation of state
\begin{equation}
P=-\frac{kT}{\vo}\ln(1-\rho \vo),
\end{equation}
exhibiting logarithmic divergence of the pressure

In the case of discrete bins, $v_{ex}^0=1, \rho=\xi$, in which case the energy \eqref{eqLinearEnergy} reduces to a simple Flory-Huggins-type entropy of mixing \cite{flory1942thermodynamics,huggins1941solutions}. In this case as the energy is symmetric under inversion $\rho\mapsto 1-\rho$, the theory of ``particles" and the theory of ``holes" are equivalent. 

\subsubsection{Uncorrelated systems and the Onsager approximation}\label{subsubsecOnsager}

In this section the state space $\Gamma$ will be general, but we have in mind the case of a positional coordinate, potentially with internal degrees of freedom such as orientation or conformation. The term ``volume" will refer to the measure on $\Gamma$. We presume that particles are uncorrelated, which we express mathematically as the approximation
\begin{equation}
{P_N(\bar{q})\approx \frac{1}{|\Gamma|^N}=\text{Un}_N(\bar{q})},
\end{equation} where $\Gamma$ is the generalised state space and $\text{Un}_N$ denotes the uniform distribution on $N$ particles in the state space $\Gamma$. Thus we may approximate 
\begin{equation}\label{eqUncorrelatedAssumption}\begin{split}
&V_f(N,V) \\
&\approx \left\langle \frac{1}{|\Gamma|}\int_{\Gamma}\exp\left(-\frac{1}{kT}\sum\limits_{i=1}^{N-1} U(q,q_i)\right)\,dq\right\rangle_{\text{Un}_{N-1}}.
\end{split}
\end{equation}
As we only consider steric interactions, we may employ a simple probabilistic argument. If the average excluded volume of a single particle in $\Gamma$ is then $v^0_{ex}$, then the probability of not intersecting a particular particle is $1-\frac{v^0_{ex}}{|\Gamma|}$, where we assume $\vo$ to be independent of the sample size. We may thus write the probability of a probe particle not intersecting {\it any} particle as 
\begin{equation}\label{eqVexOnsager}
\left(1-\frac{v_{ex}^0}{|\Gamma|}\right)^N=\left(1-\frac{v^0_{ex}\rho}{N}\right)^N,
\end{equation}
as all the particles are independently distributed.\footnote{This argument may be readily generalised in a straightforward manner to soft interactions with sufficient decay at a large distance.} This gives the cavity volume fraction in the thermodynamic limit as $N,|\Gamma|\to\infty$ as 
\begin{equation}
\mathcal{V}_f(\rho)=\exp(-v^0_{ex}\rho).
\end{equation}
We may substitute this into the free energy density equation \eqref{eqDefEnergy} to obtain 
\begin{equation}\begin{split}
\frac{1}{kT}\mf (\rho)=&\rho\ln\rho-\rho - \int_0^\rho \ln\left(\exp(-v^0_{ex}x)\right)\,dx\\
=&\rho\ln\rho-\rho +  v^0_{ex}\int_0^\rho x\,dx\\
=&\rho\ln\rho-\rho + \frac{v^0_{ex}}{2}\rho^2,
\end{split}
\end{equation}
which readily obtains the equation of state
\begin{equation}
\frac{1}{kT}P=\rho+\frac{\vo}{2}\rho^2.
\end{equation}

This produces Onsager's acclaimed free energy for a single species isotropic liquid \cite{onsager1949effects}, and correctly reclaims the second virial coefficient. Perhaps curiously, the factor of $\frac{1}{2}$ before the excluded volume does not arise as a traditional ad-hoc tool to avoiding double counting, but in a much more indirect manner via the integration step. Furthermore, this approach does not use an explicit mean-field, as is usually tradition in justifications of Onsager (see for example \cite{palffy2017onsager}, and Onsager's original work), but instead uses a purely geometric ansatz on the systems considered. While we do not attempt to estimate the error, it is immediate that the error of the approximation is precisely controlled by the validity of the approximation $P_N\approx \text{Un}_N$, and in this sense we may view Onsager as an uncorrelated limit of the system, rather than a dilute limit, although the two interpretations are not independent. 

\subsection{A fluctuating lattice approximation}\label{subsecLocalLattice}

As exact solutions are generally unavailable, and detailed numerical simulations are expensive and noisy, here we propose an alternative method of approximating the cavity volume, and thus the free energy and equation of state, that is efficient to compute and based on capturing the salient aspects of the real physical system. The key idea is to presume that the cavity volume behaves as if the local environment of each particle were a lattice, although the lattice parameters may be subject to spatial fluctuations according to a probability distribution. As such we will refer to the model as the {\it fluctuating lattice model}. In particular, this contrasts to cell theories which approximate the entire system by a single lattice, defined globally.

Consider a system of identical hard particles where the densest packing of $N$ particles forms a known lattice of volume $Nv_c$, where $v_c$ is the volume of the Voronoi cell in this lattice. If a particular configuration has total volume $V$, we define the {\it excess} volume to be $V-Nv_c$. We then propose a method for approximating the cavity volume fraction according to two assumptions:
\begin{itemize}
\item We can attribute to each particle an individual excess volume $v_e$, such that $N\langle v_e\rangle = V-Nv_c$, and $v_e$ corresponds to the amount of excess volume living within the particle's Voronoi cell. $v_e$ is a quantity that is exchanged freely and reversibly between particles as the system evolves.
\item Given a particular particle with excess volume $v_e$, its local environment is well approximated by a uniform dilation of the densest packed lattice that gives the same excess volume per particle. 
\end{itemize}
The first assumption implies that $v_e$ should be distributed according to a Boltzmann distribution (see \cite[Section 1]{pitaevskii2012physical}), so $\mathbb{P}(v_e)=\nu \exp(-\nu v_e)$, with $\langle v_e\rangle_{\mathbb{P}}=\nu^{-1}$. As the total excess volume is $V-Nv_c$, $\nu=\left(\frac{V}{N}-v_c\right)^{-1}=\frac{\rho^*\rho}{\rho^*-\rho}$ where $\rho=\frac{N}{V}$ is the number density and $\rho^*=\frac{1}{v_c}$ is the number density at the densest packing. 

The true distribution of Voronoi cell sizes in hard particle systems has been investigated previously by Senthil Kumar and Kumaran \cite{senthil2005voronoi} and empirically shown to follow a 2- or 3-parameter $\Gamma$ distribution. Within their work they also review the case for Poisson distributed points (corresponding to the dilute limit), showing there are still conflicts within the literature even in this simpler case. By using the more naive Boltzmann distribution, we give an explicit and analytic form for the probability distribution function $\mathbb{P}$, and capture salient aspects of the problem. 

The second assumption means that (locally) we have a dimensionless lattice parameter $a>1$, so that the local environment of a particle resembles a dilation of the densest lattice by factor $a$. We can relate $a$ to $v_c,v_e$ as $v_e+v_c=a^dv_c$ where $d$ is the dimension of space, which can be rearranged to give $a=\sqrt[d]{\frac{v_e}{v_c}+1}$.

Given a lattice described by spacing $a$, we denote $v_f(a)$ to be the amount of cavity volume per Voronoi cell. We may use $v_f$ to approximate the cavity volume fraction of the entire system as 
\begin{equation}\begin{split}\label{eqFluctuatingLatticeApprox}
&V_f(N,V)\\
\approx &\frac{N}{V}\int_0^\infty v_f\left(a(v_e)\right)\mathbb{P}(v_e)\,dv_e\\
=&\frac{\rho^*\rho^2}{\rho^*-\rho}\int_0^\infty v_f\left(\sqrt[d]{\frac{v_e}{v_c}+1}\right)e^{-\frac{\rho^*\rho}{\rho^*-\rho}v_e}\,dv_e
\end{split}
\end{equation}
This equation is in principle applicable to general systems, but from here we limit ourselves to hard spheres in $\mathbb{R}^d$. Before doing any numerical computations, we make a few qualitative statements about the constitutive equation.

It is immediate that for $0<\rho<\rho^*$ the integral in \eqref{eqFluctuatingLatticeApprox} is an analytic function of $\rho$, and consequently the free energy will be smooth over its entire domain. Any discontinuity in the free energy or its derivatives would necessarily have to be accompanied by corresponding discontinuities in the map $\rho\mapsto \mathbb{P}$. This statement is in accordance with the work of Senthil Kumar and Kumaran \cite{senthil2005voronoi}, which shows a sharp transition in the distribution of Voronoi cell sizes as the freezing transition is reached. Phase separation is more complicated to describe, and would have to arise from non-monotonicity of $\mathcal{V}_f$, leading to non-convexity in $\mathcal{F}$. The map $\rho\to\frac{\rho^*\rho^2}{\rho^*-\rho}\exp\left(-\frac{\rho^*\rho}{\rho^*-\rho}v_e\right)$ is bell-shaped for any $v_e>0$, which makes it unclear if $\mathcal{V}_f$ is monotonic in general, and consequently unclear if we may rule out phase separation.

Different constitutive equations on $v_f$ and $\mathbb{P}$ may be employed in \eqref{eqFluctuatingLatticeApprox} to provide alternative models. The simplest such change would be to permit different lattice structures. Employing a broader family of probability distributions, perhaps based on theoretical approximations or fitting to numerical experiments, would open other avenues for investigation as well. For the sake of this work however we will limit ourselves to the simplest possible assumptions that can provide tangible results. As we demonstrate in the sequel, these incredibly coarse assumptions are sufficient to describe the equation of state both in the high-density and dilute regimes, within a single framework. This puts it at an advantage over theories such as the virial expansion, only valid in dilute regimes, and cell theories, which are only valid in dense regimes. 

In particular, we can estimate the leading order behaviour in these extreme regimes, with the calculations deferred to \Cref{secLimitingFL}. We find that 
\begin{equation}\begin{split}
\mathcal{V}_f \approx & 1-2^d\rho v_0,
\end{split}
\end{equation}
as $\rho\to 0$, in accordance with the virial expansion. Futhermore, if there exists some $v_1>0$ such that for $v_e<v_1$, the cavity volume is zero, then for every $\epsilon>0$ there exist constants $c_\epsilon,C_\epsilon>0$ so that 
\begin{equation}\label{eqDenseScaleCav}
c_\epsilon \nu \exp(-\nu(1+\epsilon)v_1)<\mathcal{V}_f(\rho)<C_\epsilon \nu\exp(-\nu(1-\epsilon)v_1)
\end{equation}
as $\rho\to\rho^*$, where $\nu=\frac{\rho^*\rho}{\rho^*-\rho}$ as before. In particular, we may estimate the free energy and pressure as $\rho\to\rho^*$ by
\begin{equation}
-c_1\ln(\rho^*-\rho)\leq \mf(\rho)\leq -c_2\ln (\rho^*-\rho),
\end{equation}
\begin{equation}
\frac{c_3}{\rho^*-\rho}<\frac{P}{kT}<\frac{c_4}{\rho^*-\rho},
\end{equation}
for appropriate positive constants $c_i$.

\subsubsection{1D Tonks gas}
We will consider the Tonks gas as it is a simple enough system that the fluctuating lattice approximation may be calculated exactly and explicitly, as well as compared to the exact free energy density and equation of state, as in \Cref{subsubsecTonks}. We have that $v_c=2r$, and $v_f(a)=\max(0,a-2r)$. Thus the cavity volume fraction can readily be computed with a change of variables $x=v_e-2r$ as
\begin{equation}
\begin{split}
\mathcal{V}_f=& \rho\int_0^\infty \max(0,v_e-2r)\nu \exp(-\nu v_e)\,dv_e\\
=&\rho\int_{2r}^\infty \max(0,v_e-2r)\nu \exp(-\nu v_e)\,dv_e\\
=& \rho \int_0^\infty x\nu \exp(-\nu (x+2r)\,dx\\
=& \rho\exp(-2r\nu)\int_0^\infty \nu x \exp(-\nu x)\,dx\\
=& \frac{\rho\exp(-2r\nu)}{\nu}=\frac{\rho^*-\rho}{\rho^*}\exp\left(-2r\frac{\rho\rho^*}{\rho^*-\rho}\right).
\end{split}
\end{equation}
Introducing $\sigma=2r=\frac{1}{\rho^*}$ to make explicit the volume of the particles, this can be written as 
\begin{equation}
\mathcal{V}_f=(1-\rho\sigma)\exp\left(-\frac{\rho\sigma}{1-\rho\sigma}\right),
\end{equation}
which is the exact result in one dimension as seen in \Cref{subsubsecTonks}. Therefore any and all derived quantities (pressure, free energy) will also be correct using this approach. This is however a truly exceptional situation for the one-dimensional system. In a Tonks gas, our assumption of simple pairwise exchange of excess volume is more accurate, as the excess volume can be exchanged between next-nearest neighbours via translating the particle between them. In higher dimensional systems where the geometry of a cavity is no longer trivial, translations of a single particle would be expected to redistribute the excess volume in a highly complicated way between both its various neighbours and itself.

\subsubsection{2D Hexagonal lattice}
\begin{strip}
\begin{minipage}{\textwidth}
\begin{center}

\begin{figure}[H]\begin{center}
\begin{subfigure}[t]{0.45\textwidth}
\begin{center}
\includegraphics[width=0.9\textwidth]{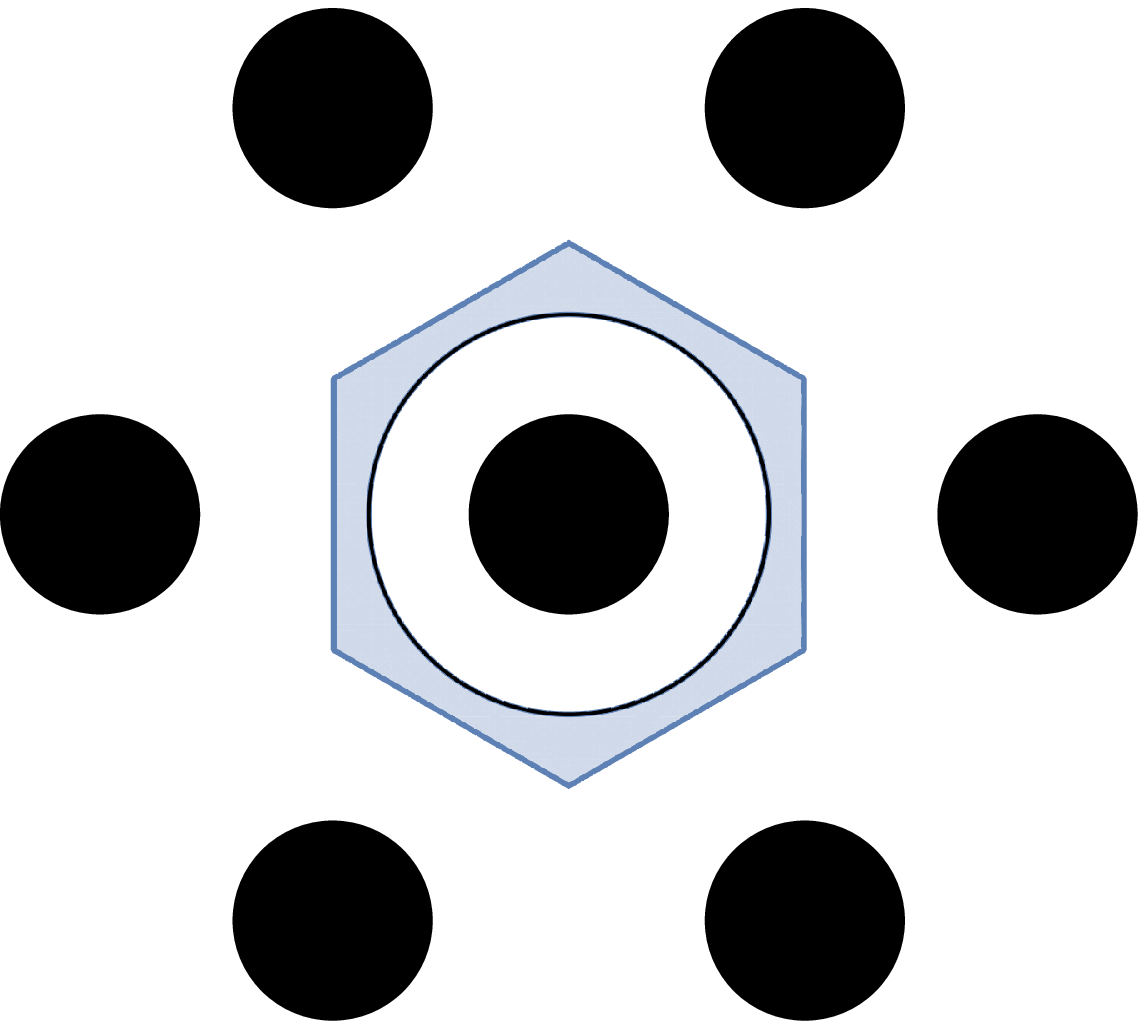}
\caption{The cavity volume that remains in a relatively dilute hexagonal lattice ($a=4.2$)
}
\end{center}
\end{subfigure}\hspace{0.05\textwidth}
\begin{subfigure}[t]{0.45\textwidth}
\begin{center}
\includegraphics[width=0.9\textwidth]{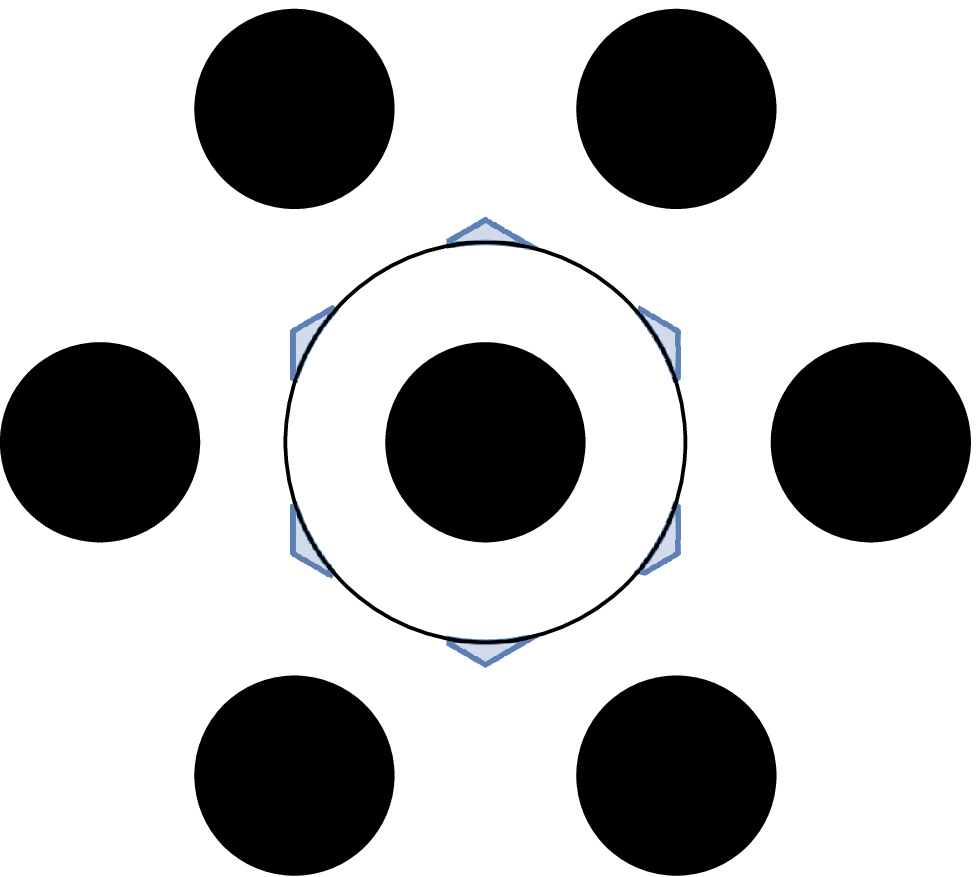}
\caption{The cavity volume that remains in a denser hexagonal lattice ($a=3.85$)
}
\end{center}
\end{subfigure}
\caption{
The figures above show the cavity volume within a single Voronoi cell. The black circles correspond to the particles in the system, the white circle surrounding the central particle is its exclusion region, and the blue region surrounding it is the cavity volume within the Voronoi cell of the central particle. 
}\end{center}
\end{figure}
\end{center}
\end{minipage}
\end{strip}

In the case of an hexagonal lattice, we can explicitly find the cavity volume function, which is done in \Cref{subsecHEXCalcs}. We evaluate the resulting equation of state by numerically integrating \eqref{eqFluctuatingLatticeApprox} to obtain $\mathcal{V}_f$, and then performing a second numerical integration in the equation of state \eqref{eqEOSUniversal}. This is then compared with the Monte Carlo simulation data of Kolafa and Rottner \cite{kolafa2006simulation} in the fluid regime, and Alder, Hoover and Young \cite{alder1968studies} in the solid regime in \Cref{figEOS2DCompare}. We constrast our results with the two-dimensional analogue of the Carnahan-Starling equation of state, which gives the compressibility factor in terms of the packing fraction $\eta$ as $Z=(1-\eta)^2$, which is expected to be accurate before the freezing transition at around $\eta\approx 0.706$. Furthermore, we make comparison with the leaky cell theory for an hexagonal lattice \cite{fai2020leaky}, which is an extension of the classical cell theory \cite{buehler1951free}, to dilute regimes. The cell theory approaches are expected to be better approximations in dense regimes. We see good agreement with the fluid regime up to moderate densities, with accuracy becoming worse as we approach the freezing transition. However once  we enter the solid regime, we see very good agreement between the fluctuating lattice model and the Monte Carlo data. We include the relative error between the measurements of Alder, Hoover and Young and the fluctuating lattice model in \Cref{figRatios2D}, and we see that the relative accuracy is improving towards the dense limit, and furthermore is within 5\% beyond $\eta=0.8$. As the quality of approximation of the equation of state is directly related to the quality of approximation of the cavity volume, we show in \Cref{figDilute2D} a comparison of the cavity volume against an ad-hoc Monte Carlo experiment up to modest densities ($\eta=0.6$, details in \Cref{subsecMC}), in which we see a consistent over-estimate. Due to the expectation of vanishingly small cavity volume in higher-density regimes, where incredibly large systems and long runs would be required to obtain stable results, we do not probe the high-density regime in our Monte Carlo methods.

Given our asumptotic result \eqref{eqDenseScaleCav} in the dense regime, we expect that $\mathcal{V}_f$ should behave as $c_1\exp\left(-\frac{c_2}{\rho^*-\rho}\right)$ as $\rho\to\rho^*$ for some constants $c_1,c_2$. We perform a linear least-squares regression of $\frac{-1}{\ln \mathcal{V}_f}$ versus $\eta^*-\eta$. We see a good linear fit, as shown in \Cref{fig2DLinearFit}, so that $\frac{-1}{\ln \mathcal{V}_f}\approx 0.498(\eta^*-\eta)$, which would give the leading order approximation as $\mathcal{V}_f\approx c e^{-\frac{2.08}{\eta^*-\eta}}$.

\begin{strip}
\begin{minipage}{\textwidth}
\begin{figure}[H]
\begin{center}
\includegraphics[width=0.6\textwidth]{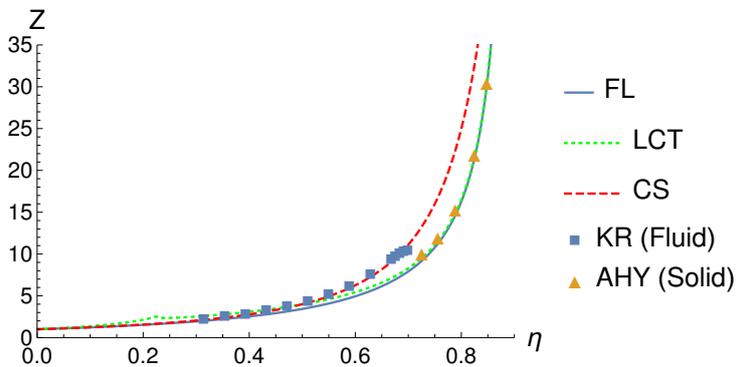}
\caption{Comparison of the data of Kolafa and Rotter (KR), and Alder, Hoover and Young (AHY) against the obtained equation of state for the fluctuating lattice model (FL) for the two-dimensional hard disk system. The two-dimensional Carnahan-Starling (CS) and Leaky Cell Theory (LCT) equations of state are included for comparison.}
\label{figEOS2DCompare}
\end{center}
\end{figure}
\end{minipage}\vspace{-1.5cm}
\end{strip}

\begin{strip}
\begin{minipage}{\textwidth}
\begin{figure}[H]
\begin{center}
\begin{subfigure}[t]{0.31\textwidth}
\begin{center}
\includegraphics[width=\textwidth]{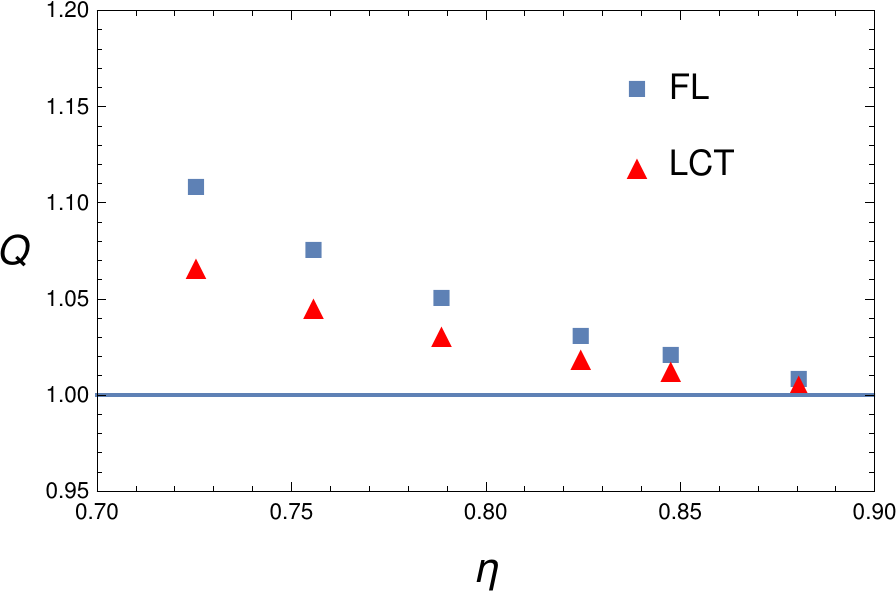}
\caption{Quotients $Q=P_{HAY}/P$ with the pressure $P$ obtained from the Leaky Cell Theory (LCT) and Fluctuating Lattice (FL) against the numerical data $P_{HAY}$ of Alder, Hoover and Young in the dense regime.}\label{figRatios2D}
\end{center}
\label{figEOS2DRatio}
\end{subfigure}
\hspace{0.025\textwidth}
\begin{subfigure}[t]{0.31\textwidth}
\begin{center}
\includegraphics[width=\textwidth]{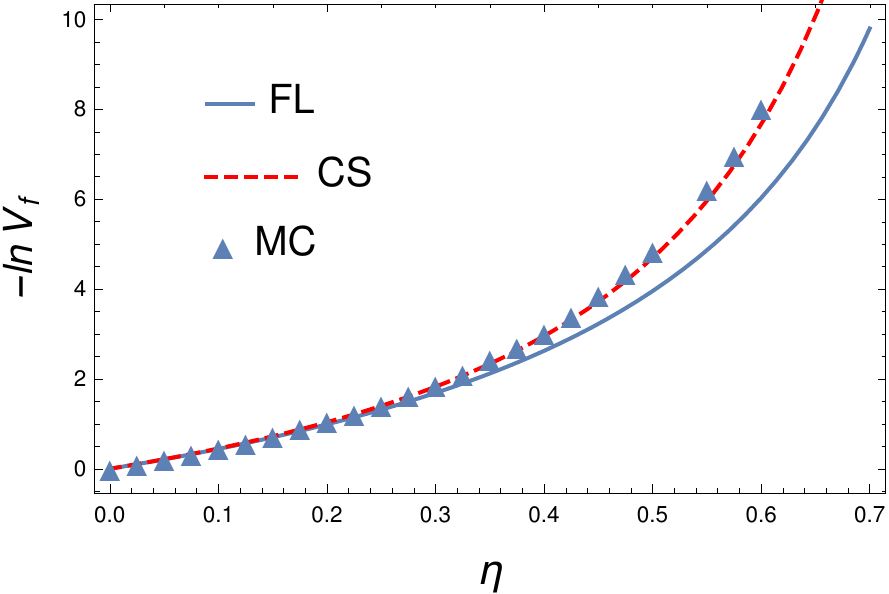}
\caption{Comparison of the cavity volume fraction from the fluctuating lattice approximation with ad-hoc Monte Carlo runs, and the inferred cavity volume of the 2-dimensional Carnahan-Starling, obtained by \eqref{eqInvertEOS}, using a logarithmic scale.}\label{figDilute2D}
\end{center}
\end{subfigure}\hspace{0.025\textwidth}
\begin{subfigure}[t]{0.295\textwidth}
\begin{center}
\includegraphics[width=\textwidth]{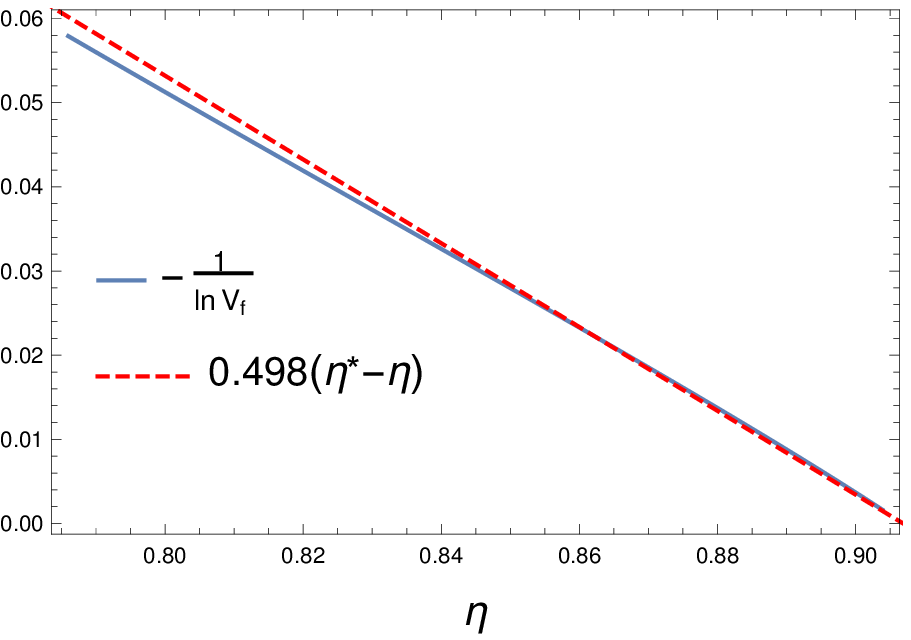}
\caption{Linear fit of $-(\ln\mathcal{V}_f)^{-1}$ against $\alpha(\eta^*-\eta)$ in the dense regime with optimal constant found to be $\alpha\approx 0.498$}
\label{fig2DLinearFit}
\end{center}
\end{subfigure}
\caption{Further comparisons of the fluctuating lattice model for the two-dimensional hard disk system}
\label{fig2dComparisonsMisc}
\end{center}
\end{figure}
\end{minipage}
\end{strip}

\subsubsection{3D FCC lattice}
We now consider the case of a 3-dimensional FCC lattice. We defer calculations of the cavity volume function $v_f$ to \Cref{subsecFCCCalcs}. We compare the packing fraction $\eta=\rho\frac{4\pi}{3}r^3$ against the compressibility $Z=\frac{P}{\rho kT}$ in \Cref{fig3DEOSComparison}. Furthermore, as in the two-dimensional case,  we compare the obtained equation of state against the classical Carnahan-Starling equation of state given in terms of the packing density as $Z=
\frac{1+\eta+\eta^2-\eta^3}{(1-\eta)^3}$, the Leaky Cell Theory for an FCC lattice, and Monte Carlo data, in this case from Wu and Sadus \cite{wu2005hard}. \pagebreak

\begin{strip}
\begin{minipage}{\textwidth}
\begin{figure}[H]
\begin{subfigure}[t]{0.45\textwidth}
\begin{center}
\includegraphics[width=\textwidth]{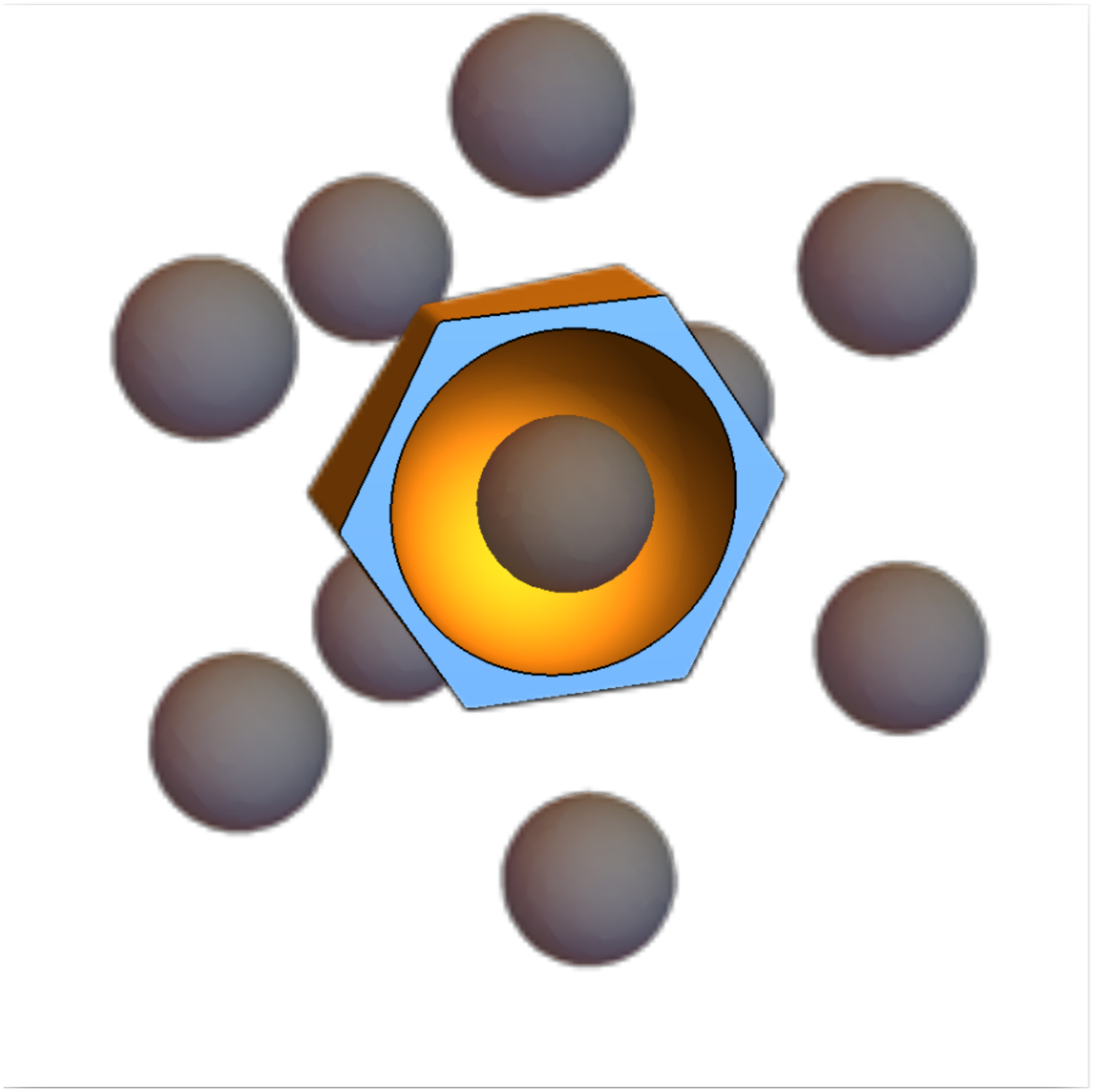}
\caption{The cavity volume that remains in a relatively dilute FCC lattice ($a=4.5$)
}
\end{center}
\end{subfigure}
\hspace{0.05\textwidth}
\begin{subfigure}[t]{0.45\textwidth}
\begin{center}
\includegraphics[width=\textwidth]{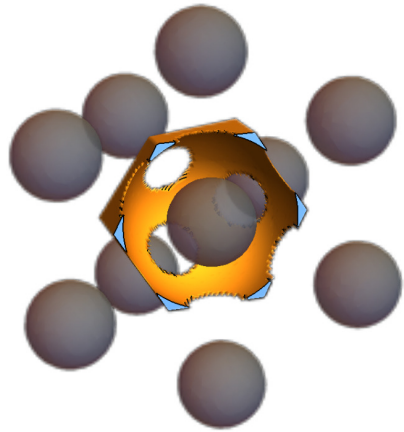}
\caption{The cavity volume that remains in a denser FCC lattice ($a=3.85$)
}
\end{center}
\end{subfigure}
\caption{
The figures above show the cavity volume within a single Voronoi cell. Only the part of the cavity volume with $x_3>0$ is shown for visibility. The black spheres correspond to the particles in the system. The region around the central particle is the cavity volume within the Voronoi cell of the central particle. 
}
\end{figure}
\end{minipage}
\end{strip}

\begin{strip}
\begin{minipage}{\textwidth}
\begin{figure}[H]
\begin{center}
\includegraphics[width=0.6\textwidth]{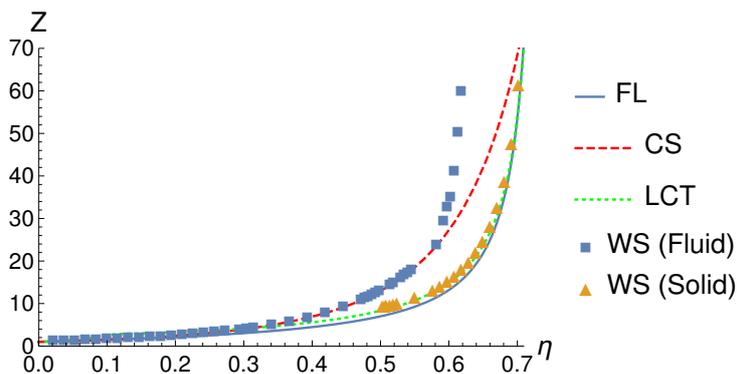}
\caption{Comparison of the fluctuating lattice (FL) equation of state for three dimensional hard spheres, with the commonly used Carnahan-Starling (CS), the Leaky Cell Theory (LCT), and comparison to Monte Carlo simulation data of Wu and Sadus (WS), with the fluid branch and solid branch of their data separated for comparison. Note that for $\eta$ greater than $\approx 0.502$, the fluid branch is only metastable.}
\label{fig3DEOSComparison}
\end{center}
\end{figure}
\end{minipage}
\end{strip}

As expected from our analysis, there is generally good agreement with the Monte-Carlo data and the Carnahan-Starling equations of state in the dilute regime. However in intermediate, fluid, regimes we start to see stronger disagreement. We see that for the entirety of the solid branch we have a modest agreement with the fluctuating lattice equation of state. In \Cref{figRatiosEOS}, we see a monotonic increase in relative accuracy of the fluctuating lattice equation of state in the solid phase. Furthermore, we see that the fluctating lattice equation of state always provides an underestimate of the compressibility factor. Similarly to the two-dimensional case, we perform a linear fit of $-(\ln\mathcal{V}_f)^{-1}$ against $\alpha(\eta^*-\eta)$, shown in \Cref{fig3DLinearFit}, which shows good agreement with $\alpha =0.413$. This implies an asymptotic approximation $\mathcal{V}_f\approx c\exp\left(-\frac{2.32}{\eta^*-\eta}\right)$ in the dense limit.  We can directly compare the calculated cavity volume against Monte Carlo data in intermediate regimes, which we show in \Cref{fig3DCavityVol}.
\pagebreak

\begin{strip}
\begin{minipage}{\textwidth}
\begin{figure}[H]
\begin{center}
\begin{subfigure}[t]{0.3\textwidth}
\begin{center}
\includegraphics[width=0.95\textwidth]{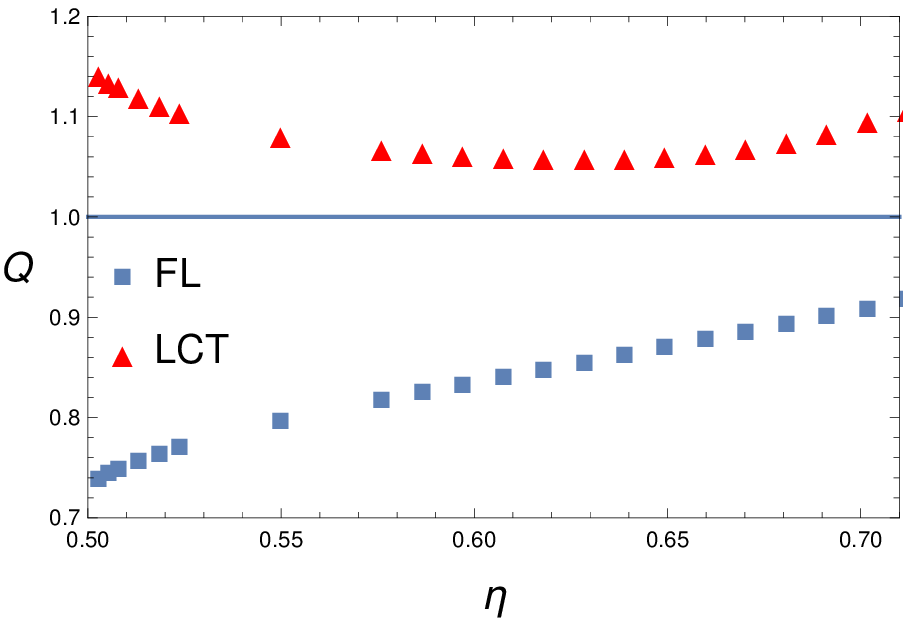}
\caption{The quotients $Q=P_{WS}/P$ of the Leaky Cell Theory (LCT) and Fluctuating Lattice (FL) pressures $P$ with the data of Wu and Sadus $P_{WS}$ on the solid branch.
}\label{figRatiosEOS}
\end{center}
\end{subfigure}\hspace{0.025\textwidth}
\begin{subfigure}[t]{0.3\textwidth}
\begin{center}
\includegraphics[width=0.975\textwidth]{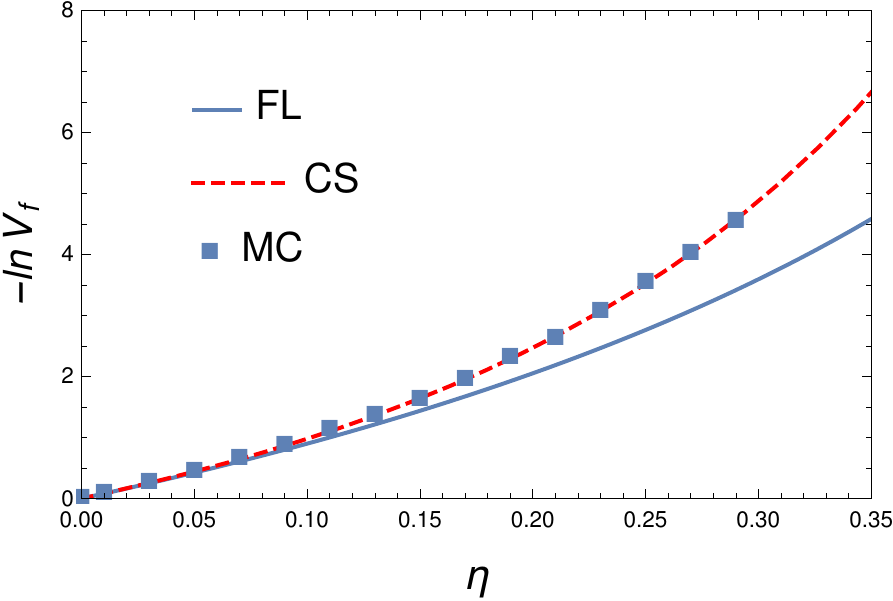}
\caption{Comparison of the fluctating lattice prediction of cavity volume fraction against Monte Carlo simulation, and the inferred cavity volume of the 3-dimensional Carnahan-Starling, obtained by \eqref{eqInvertEOS}, using a logarithmic scale.}
\label{fig3DCavityVol}
\end{center}
\end{subfigure}\hspace{0.025\textwidth}
\begin{subfigure}[t]{0.3\textwidth}
\begin{center}
\includegraphics[width=0.925\textwidth]{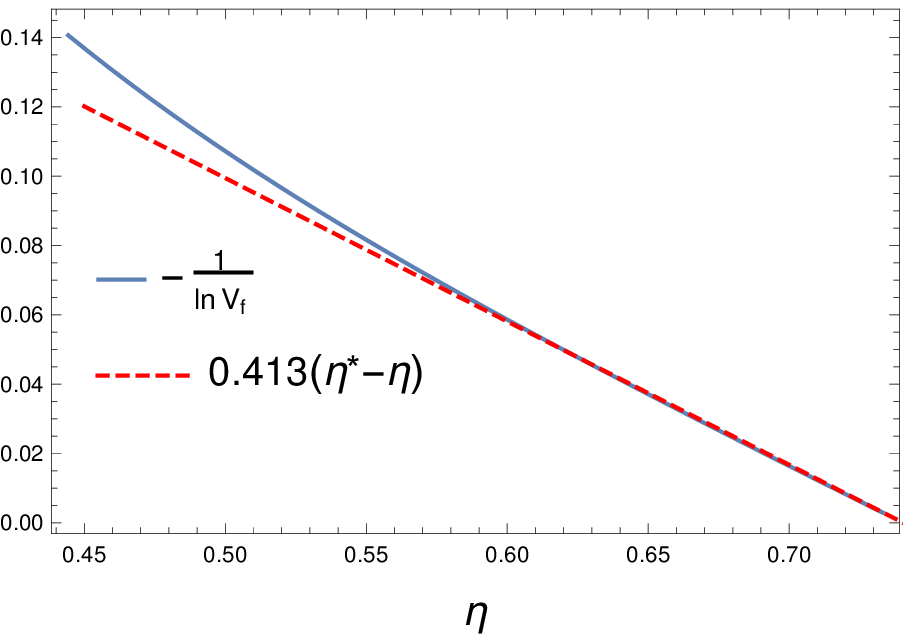}
\caption{Linear fit of $-\ln(\mathcal{V}_f)^{-1}$ against $\alpha(\eta^*-\eta)$ in the dense regime with optimal constant found to be $\alpha\approx 0.413$}
\label{fig3DLinearFit}
\end{center}
\end{subfigure}
\caption{ Further comparisons of the fluctuating lattice model for the three-dimensional hard sphere system}
\end{center}
\end{figure}
\end{minipage}
\end{strip}

\section{Concluding remarks}

Within our work we have used the notion of cavity volume to provide expressions for the free energy and equation of state in single-species systems by performing a type of thermodynamic integration with respect to number density from the vacuum state. The framework we have used, owing to its generality, permits us to reclaim known results such as the Onsager approximation and exact solution of the Tonks gas in one dimension. Our strategy offers new expressions for the virial coefficients given in terms of the cavity volume in the dilute regime, showing that if the cavity volumes of systems of up to $n-1$ particles are known, in principle, we may evaluate the reduced virial coefficient $\text{b}_n$. 

This has strengthened our confidence in the theoretical setting envisioned here. A number of novel conclusions come from our proposed approach. We systematised a heuristic expression for the cavity volume, which had already proven predictive in \cite{nascimento2017density}. We profited from the generality we can afford by identifying a constitutive function $\lambda$ that we have newly interpreted as an effective correction for overlapping exclusion zones between interacting particles. 

In addition, we proposed a {\it fluctuating lattice model}, which is obtained by making simplifying assumptions in order to calculate the cavity volume fraction in hard particle systems. The model has shown that with a small number of assumptions, we can capture the salient aspects of the equation of state for hard rods, disks and spheres, in different space dimensions, in both the dilute and dense regimes. This offers an advantage over the virial expansion and Carnahan-Starling equation of state, which are based on series expansions in the dilute regime and therefore cannot be expected to provide accurate results beyond a phase transition; as well as advantages over the cell theories, which perform well near dense packings but however fail to capture the correct behaviour in dilute regimes.

The main limitation of the fluctuating lattice model as presented in our work is that it does not capture phase transitions, presumably as a consequence of the independence of the cell cavity volume function $v_f$ and the density, which ignores qualitative changes in the geometry of the Voronoi cell as the density varies. Similarly, the assumptions that lead to a Boltzmann distribution of the excess cavity volume, while being convenient for the arithmetic and analysis of the free energy, contradict known results even in the dilute regime (see \cite{ferenc2007size,senthil2005voronoi}). Improving the structural assumptions on the distribution of excess cavity volume and the function $v_f$ remains an avenue for future investigation, whereby permitting competition between different lattice structures could lead to a model capable of describing phase transitions. Furthermore, while only hard interactions have been considered in the fluctuating lattice model within this work, the same methodology would be extendable to long-range, soft interactions, which may or may not exist in tandem with hard-core repulsion at short range. As well as broadening the nature of particle interactions, we believe this framework will be extendable to systems with more complex structures, namely anisotropic and multi-species systems. We have in mind liquid crystalline systems as a future avenue, where suitable adjustments to the model could provide a classical density functional theory for spatially inhomogeneous liquid crystalline systems, similar to the approach as seen in \cite{nascimento2017density}.

\section{Acknowledgments}
T.G.F. acknowledges support from National Science Foundation grant DMS-1913093. P.P-M. acknowledges support from the Office of Naval Research (ONR N00014-18-1-2624). J.M.T. has been
partially supported by the Basque Government through the BERC 2018-2021 program; and by Spanish Ministry of Economy and Competitiveness MINECO through BCAM Severo Ochoa excellence accreditation SEV-2017-0718 and through project MTM2017-82184-R funded by (AEI/FEDER, UE) and acronym “DESFLU". Finally, we wish to acknowledge the support of the Institute for Mathematics and its Applications (IMA), where this work was initiated during the 2018 program on “Multiscale Mathematics and Computing in Science and Engineering”.

\appendix

\section{MC simulations}\label{subsecMC}

Approximations and constitutive equations of the excluded volume can tell us certain information relating to the system, but ultimately in 2 or more dimensions we are constrained to numerical methods if we wish to obtain quantitatively accurate data for the excluded volume of even modest density systems. For this we perform a simple Monte Carlo study to provide results for comparison. The methodology is as follows:
\begin{enumerate}
\item Equilibrate a configuration by moving each particle at least 50 successful Monte Carlo steps.
\item Estimate the cavity volume by testing if 150 randomly chosen points lie in any exclusion regions.
\item Generate a new realisation by further Monte Carlo methods on the previous realisation, until all particles have been successfully moved at least 10 times. 
\item Go back to (2), unless the error of the logarithm of the cavity volume is sufficiently small. 
\end{enumerate} 
We consider the error of the logarithm as this is the quantity appearing in the energy, and this means that in denser systems a significantly larger number of runs is required as $\mathcal{V}_f$ is close to zero, meaning small fluctuations lead to large fluctuations in $-\ln \mathcal{V}_f$. We estimate the error of the logarithm as follows. Given $n$ test points and $s$ points in the cavity space, we estimate a 95\% confidence interval using the Wilson score interval, 
\begin{equation}
p=\frac{2s+z^2}{2n+2z^2}\pm \frac{z}{n+z^2}\sqrt{\frac{s(n-s)}{n}+\frac{z^2}{4}}.
\end{equation}
Here $z=1.96$ corresponds to the $95\%$ confidence interval, $s$ is the number of successful trials and $n$ is the total number of trials. 
Then we estimate the error of the logarithm as the width of the logarithm of this interval.

\section{Limiting behaviour in the fluctuating lattice model}
\label{secLimitingFL}
\subsection{Dilute systems}
Let the local lattice spacing be $a$. We note that if $a=1$ we are at the densest packing and no particles overlap, and are thus spaced at least by twice their radius $r$. When $a>2$, all particles must be at distance at least $4r$, meaning no exclusion regions overlap. In particular, this means that $v_f(a)=a^dv_c-2^dv_0$, where $v_0$ is the volume of one particle and $2^dv_0$ is the pairwise excluded volume. This can be simplified as $v_f(a(v_e))=v_e+v_c-2^dv_0$. This means we can write the cavity volume fraction, writing $\nu = \frac{\rho^*\rho}{\rho^*-\rho}$ and $v_1$ to be the solution of $a(v_1)=2$ as 
\begin{equation}\begin{split}
&\mathcal{V}_f\\
=&\rho\left(\int_0^{v_1}\nu \exp(-\nu v_e)\left(v_f(a(v_e))-a(v_e)^dv_c+2^dv_0\right)\,dv_e\right.\\
&\hspace{0.5cm}\left.+\int_0^\infty (a(v_e)^dv_c-2^dv_0)\nu\exp(-\nu v_e)\,dv_e\right)\\
\approx & \rho\left(c\nu+\int_0^\infty (v_c-2^dv_0+v_e)\nu\exp(-\nu v_e)\,dv_e\right)\\
= & \rho\left(c\nu +v_c-2^dv_0+\frac{\rho^*-\rho}{\rho^*\rho}\right),
\end{split}
\end{equation}
where $c$ is an order 1 quantity as $\rho\to 0$, or equivalently $\nu \to 0$. Thus in the dilute regime $\rho\approx 0$, we have that the leading order contribution to $\mathcal{V}_f$ is 
\begin{equation}\begin{split}
\mathcal{V}_f\approx &\rho v_c-\rho 2^dv_0+1-\frac{\rho}{\rho^*}\\
=& 1-2^d\rho v_0,
\end{split}
\end{equation}
recalling $v_c=\frac{1}{\rho^*}$. We thus reclaim the expected first order behaviour of the cavity volume. It should be noted that this argument depends only on the fact that $\mathbb{P}(v_e)\to 0$ for $0<v_e<v_1$ as $\rho \to 0$, and would extend to more general systems without issue if the constitutive probability distributions $\mathbb{P}$ satisfied this. 

\subsection{Dense systems}

Let $\rho\approx \rho^*$. Assume that $v_f(a(v_e))=0$ for $v_e<v_2$, and $v_2>0$. That is, in a nearly-dense packing there is no cavity volume, which is certainly reasonable for spheres. Then for any fixed $1>\delta>0$, $\nu$ sufficiently large, and $v_e>v_2$, we have that $\nu v_e> (\delta \nu -1)v_e> \delta \nu v_2-v_e$. Then this means that 
\begin{equation}
\begin{split}
&\mathcal{V}_f\\
=&\int_{v_2}^\infty \rho v_f\left(\sqrt[d]{\frac{v_e}{v_c}+1}\right)\nu\exp(-\nu v_e)\,dv_e\\
\leq &\nu\exp(-\delta\nu v_2) \int_{v_2}^\infty \rho v_f\left(\sqrt[d]{\frac{v_e}{v_c}+1}\right)\exp(- v_e)\,dv_e.
\end{split}
\end{equation}
We note that the integral is finite and independent of $\nu$. Therefore we have a bound 
\begin{equation}
\mathcal{V}_f\leq c_\delta\nu\exp(-\delta v_2\nu)=\frac{c_\delta\rho\rho^*}{\rho^*-\rho}\exp\left(-\frac{\delta\rho^*\rho}{\rho^*-\rho}v_2\right).
\end{equation}
for some $c_\delta>0$ and $0<\delta<1$, for sufficiently large $\nu$ (equivalently, $\rho$ sufficiently close to $\rho^*$).

By a similar argument, if $\gamma>1$, using that $\nu v_e < (\gamma \nu +1)v_e\leq \gamma\nu v_e+\gamma^2 v_2$ for $v_2<v_e<\gamma v_2$, 
\begin{equation}
\begin{split}
&\mathcal{V}_f\\
=&\int_{v_2}^\infty \rho v_f\left(\sqrt[d]{\frac{v_e}{v_c}+1}\right)\nu\exp(-\nu v_e)\,dv_e\\
\geq &\nu\exp(-\nu \gamma^2 v_2)\int_{v_2}^{\gamma v_2} \rho v_f\left(\sqrt[d]{\frac{v_e}{v_c}+1}\right)\exp( -  v_e)\,dv_e.
\end{split}
\end{equation}
This gives a similar scaling, that for any $\gamma>1$, for $\rho$ sufficiently close to $\rho^*$, 
\begin{equation}\begin{split}
&\mathcal{V}_f\\
\geq &c_\gamma \nu \exp(-\nu \gamma^2 v_2)\\
=&\frac{c_\gamma\rho\rho^*}{\rho-\rho^*}\exp\left(-\frac{\gamma^2\rho\rho^*}{\rho^*-\rho}v_2\right).\end{split}
\end{equation}

This means that the cavity volume tends to zero faster than any polynomial, and the upper bound is a near-classical example of a function with vanishing derivatives of all orders as $\rho\to \rho^*$, despite being a non-zero function. Written in another way, we can say that for sufficiently small $\epsilon>0$ there exists constants $c_\epsilon,C_\epsilon>0$ so that 
\begin{equation}\begin{split}
&c_\epsilon \nu \exp(-\nu(1+\epsilon)v_1)\\
&<\mathcal{V}_f(\rho)\\
&<C_\epsilon \nu\exp(-\nu(1-\epsilon)v_1).
\end{split}
\end{equation}

The argument we have presented exemplifies a key point that we have seen in the Tonks gas, that the leading cause for the loss of cavity volume is not that particular cavities are exceptionally small, but that they are exceptionally {\it rare}, in dense systems. This is quantified in the preceding argument by noting that moderately dense lattices still have zero cavity volume, and thus we may estimate the total cavity volume by the ``tail" of the distribution $\mathbb{P}$, which is rapidly decaying. 

Substituting these inequalities into \eqref{eqDefEnergy} provides the estimates
\begin{equation}
-c_1\ln(\rho^*-\rho)\leq \mf(\rho)\leq -c_2\ln (\rho^*-\rho),
\end{equation}
\begin{equation}
\frac{c_3}{\rho^*-\rho}>\frac{P}{kT}>\frac{c_4}{\rho^*-\rho}
\end{equation}
as $\rho\to \rho^*$.

\section{Cavity volume calculations}
\subsection{Cavity volume of a 2D Hexagonal lattice}
\label{subsecHEXCalcs}
As the configuration is highly symmetric it suffices to consider half a unit cell, which at spacing $a>0$ has three relevant particles sitting on vertices of an equilateral triangle of side length $2ar$, where $r$ is the particle radius.  We see there are three regimes. When $a<\sqrt{3}$ the centre point is contained in all exclusion regions and there is zero cavity volume. When $a >2$ all particles are at a distance of at least $4r$ and no exclusion zones overlap. thus it suffices to evaluate the overlaps in the intermediate regime. For $\sqrt{3}<a<2$, we must account for the overlap of the exclusion regions. We may thus compute the cavity volume within the triangle as
\begin{equation}
\frac{\sqrt{3}}{4}(2ar)^2-3\frac{1}{6} 4\pi r^2+\frac{3}{2}V_2(a).
\end{equation}
The first term corresponds to the area of the triangle, the second to the removal of the exclusion regions of the three particles, of which only a sixth is inside the triangle. The final term accounts for double overlaps, where $V_2(a)$ is the intersection area of two disks of radius $2r$ at a distance $2ar$, given explicitly by $V_2(a)=8r^2\cos^{-1}\left(\frac{a}{2}\right)-2ar^2\sqrt{4-a^2}$.  The densest packing occurs at $\rho^*=\frac{1}{2\sqrt{3}r^2}$.

\subsection{Cavity volume function of a 3D FCC lattice}
\label{subsecFCCCalcs}
Here we calculate the cavity volume function for an FCC lattice of spacing $a>1$. We perform this analysis by considering a unit cell, which contains the equivalent of four particles (three halves, and eight quarters). The unit cell can be decomposed into $N_t=8$ tetrahedra and $N_o=4$ octahedra, where each vertex is the centre of mass of a particle. We note that the following calculations also apply to an HCP structure, as its unit cell can be deconstructed into equivalent blocks. Thus it suffices to find the cavity volume in a particular tetrahedron and octohedron. We introduce the following notation. $V_2(a)$ will denote the intersection volume of two spheres of radius $2$ at distance $2a$. $V_{3}^{eq}(a)$ denotes the intersection volume of three spheres on the vertices of an equilateral triangle of side length $2a$, and $V_3^{rt}(a)$ denotes the intersection volume of three spheres on the vertices of a right-angled triangle of hypotenuse length $2\sqrt{2}a$. We needn't consider intersections of higher order, as we will have hard particles of radius $1$, and in an FCC lattice the intersection of four exclusion spheres only appears when the cavity volume is zero. We can evaluate the corresponding functions using formulae from \cite{gibson1987volume} as 

\begin{strip}
\begin{equation}
\begin{split}
V_2(a)=&\left\{\begin{array}{c l}
\frac{\pi}{12}(4-2a)^2(8+2a) & a<2\\
0 & a\geq 2
\end{array}\right.\\
V_3^{eq}(a)=&\left\{\begin{array}{c l}
\frac{2a^2}{3}\sqrt{12-4a^2}+32\tan^{-1}\left(\frac{\sqrt{12-4a^2}}{2}\right)-3a\left(8-\frac{2a^2}{3}\right)\tan^{-1}\left(\frac{\sqrt{12-4a^2}}{a}\right) & a<\sqrt{3}\\
0 & a\geq \sqrt{3}
\end{array}\right.\\
V_3^{rt}(a)=&\left\{\begin{array}{c l}
\frac{2a^2}{3}\sqrt{16-8a^2}-\frac{a\pi(24-a^2)}{3\sqrt{2}}+ \frac{32}{3}\left(\frac{\pi}{2}+2\tan^{-1}\left(\frac{\sqrt{16-8a^2}}{4}\right)\right)\\
-2a\left(8-\frac{2a^2}{3}\right)\tan^{-1}\left(\frac{\sqrt{16-8a^2}}{2a}\right) & a<\frac{2\sqrt{6}}{3}\\
0 & a\geq \frac{2\sqrt{6}}{3}
\end{array}\right.
\end{split}
\end{equation}
\end{strip}
Furthermore, we introduce $\Omega_t=\cos^{-1}\frac{23}{27},\Omega_o=4\sin^{-1}\frac{1}{3}$ as the solid angles of the vertices of the tetrahedron and octahedron, respectively, and $\alpha_t=\cos^{-1}\frac{1}{3},\alpha_0=\cos^{-1}-\frac{1}{3}$, as the corresponding dihedral angles. The volumes of the corresponding tetrahedron and octahedron respectively are $V_t=\frac{\sqrt{2}(2a)^3}{6},V_o=\frac{\sqrt{2}(2a)^3}{3}$. Then we can evaluate the cavity volumes of a single tetrahedron and octahedron, $v_f^t,v_f^o$ respectively as

\begin{strip}
\begin{equation}
\begin{split}
v_f^t(a)=&V_t-4\frac{\Omega_t}{4\pi}\frac{32\pi}{3}+6\frac{\alpha_t}{2\pi}V_2(a)-4\frac{1}{2}V_3^{eq}(a)\\
v_f^o(a)=&V_t-6\frac{\Omega_o}{4\pi}\frac{32\pi}{3}+12\frac{\alpha_o}{2\pi}V_2(a)-8\frac{1}{2}V_3^{eq}(a)-4V_3^{rt}(a)\\
\end{split}
\end{equation} 
\end{strip}
when $a>\frac{2\sqrt{3}}{6}$ and zero otherwise. These formulae arise simply from adding even-numbered overlaps and subtracting odd-numbered overlaps, and factors $\frac{\Omega_X}{4\pi}$, $\frac{\alpha_X}{2\pi}$ account for the ratio of the sphere contained within the polyhedra, with the integer coefficients count how many relevant (combinations of) spheres need to be considered. This gives the cavity volume per particle as 
\begin{equation}
v_f(a)=\frac{1}{4}\left(8v_f^t(a)+4v_f^o(a)\right),
\end{equation}
where the prefactor $\frac{1}{4}$ accounts for the fact there are four particles per unit cell, and the coefficients $8,4$ correspond to the number of tetrahedra and octahedra in each cell. 
\small
\bibliography{bibl}

\end{document}